\documentclass[useAMS,usenatbib,usegraphicx]{mn2e}

\usepackage{lscape}
\usepackage{txfonts}

\usepackage{biblio}
\usepackage{color}

\title[The role of E+A and post-starburst galaxies -- I. Models and model results]{The role of E+A and post-starburst galaxies\\I. Models and model results}
\author[M. A. Falkenberg, R. Kotulla \& U. Fritze]{M. A. Falkenberg$^{1}$\thanks{E-mail:
    atyra@astro.physik.uni-goettingen.de}, R. Kotulla$^{2}$ and U. Fritze
$^{2}$\thanks{E-mail: r.kotulla@herts.ac.uk, u.fritze@herts.ac.uk}\\
$^{1}$Institut f\"ur Astrophysik G\"ottingen, Georg-August Universit\"at G\"ottingen,
            Friedrich-Hund-Platz 1, 37077, G\"ottingen\\
$^{2}$Centre for Astrophysics Research, University of Hertfordshire, College Lane, Hatfield AL10 9AB, UK}
\begin{document}

\date{Accepted 2008 Month ??. Received 2008 Month ??; in original form 2008 Month ??}

\pagerange{\pageref{firstpage}--\pageref{lastpage}} \pubyear{2008}

\maketitle

\label{firstpage}

\begin{abstract}
Different compositions of galaxy types in the field in comparison to galaxy
clusters as described by the morphology-density relation in the local universe
is interpreted as a result of transformation processes from late- to early-type
galaxies. This interpretation is supported by the Butcher-Oemler effect.  We
investigate E+A galaxies as an intermediate state between late-type galaxies in
low density environments and early-type galaxies in high density environment to
constrain the possible transformation processes.

For this purpose we model a grid of post-starburst galaxies by inducing a burst
and/ or a halting of star formation on the normal evolution of spiral galaxies
with our galaxy evolution code GALEV.  From our models we find that the common
E+A criteria exclude a significant number of post-starburst galaxies and propose
that comparing their spectral energy distributions leads to a more sufficient
method to investigate post-starbust galaxies. We predict that a higher number of
E+A galaxies in the early universe can not be ascribed solely to a higher number
of starburst, but is a result of a lower metallicity and a higher burst strength
due to more gas content of the galaxies in the early universe. We find that even
galaxies with a normal evolution without a starburst have a H$\delta$-strong
phase at early galaxy ages.
\end{abstract}


\section{Introduction}\label{intro}

The composition of galaxy types is different in the field (=low galaxy density
environments) as compared to galaxy clusters (=high galaxy density environment).
In the local universe this e.g. is described by the morphology-density (MD)
relation \citep{Dressler1980ApJ, Oemler1974ApJ}.  The higher the projected
surface density of galaxies within an area, the higher is the fraction of
early-type galaxies and the lower is the fraction of late-type galaxies
\citep{Dressler1997}. The field is rich in spirals and irregular galaxies, while
in clusters mainly S0s, dSphs and dEs are found.  Butcher and Oemler were the
first to report a high fraction of blue galaxies in distant clusters as compared
to local ones (BO effect, see \citet[][]{Butcher1984, Butcher1978ApJ,
  Ellingson2001ApJ, Kodama2001MNRAS}. Since galaxy clusters grow by continuous
accretion of the field galaxies, this difference is not caused by different
formation histories but is due to a transformation that takes place during the
lifetime of a galaxy \citep{Couch1994ApJ, Dressler1994ApJ}. This interpretation
is supported by findings that the fraction of ellipticals remains constant in
clusters of different redshift, while the fraction of spirals increases and the
fraction of S0 galaxies correspondingly decreases with increasing redshift
\citep[see][]{Fasano2000}. When galaxies encounter a high density environment
like a galaxy cluster the star formation (SF) is suppressed causing an evolution
from late to early spectral types for these galaxies.  This shows that
environment plays a crucial role in the evolution of galaxies.  While in earlier
times it was believed that galaxy transformation takes only place in clusters
and is related to the Intra Cluster Medium (ICM), today we know that galaxy
properties like the SF activity, the \mbox{H\,{\sc i}}. content and
morphological type depend on the local galaxy density in general and the ICM
cannot be the only explanation to it.  When spectra of the blue population of
galaxies in clusters were investigated, most BO galaxy spectra showed strong
emission lines typical of blue star forming galaxies. However some of the blue
galaxies were discovered to show spectra with strong Balmer absorption lines,
but no emission lines \citep{Couch1987MNRAS}. These spectra are different from
the spectra of any normal Hubble type galaxy and were named E+A galaxies. In
many clusters, these E+A galaxies constitute a large fraction of the blue galaxy
population \citep{Barger1996MNRAS, Poggianti1996}.  Some red galaxies in
clusters were also discovered to show strong Balmer absorption lines. This
effect is called the "spectroscopic BO-effect".  X-ray observations have shown
that E+A galaxies can be found near the edges of infalling structures in
clusters located in an area between early-type galaxies in the central cluster
regions and late-type galaxies in the field and cluster outskirts
\citep{Poggianti2004ApJ}.  Investigations of \citet{Helmboldt2007MNRAS} show
that the gas content of E+A galaxies on average lies between gas-rich disk
galaxies and typical gas poor E/S0 galaxies.

Several processes have been proposed as mechanisms for the transformation of
galaxies in cluster environments. The most common ones are mergers, harassment,
gas stripping and strangulation.  \emph{Harassment} is an effect of tidal forces
caused by a series of many weak galaxy-galaxy encounters with high relative
velocities. The encountering galaxies are influenced by each other's
gravitational field. They strip off stars and gas from each other's outer
regions and weaken or even destroy their internal stabilities and disks. This
can result in a star burst and/or the termination of SF on a long timescale of
1-5 Gyr. Harassment is a theoretically motivated mechanism, which has been
observed in numerical simulations \citep[see][]{Moore1998ApJ,
  Richstone1976ApJ}. It is believed to play an important role in rich galaxy
clusters.  \emph{Gas stripping} describes interactions between the gas in the
disk of a galaxy and the hot and dense Intra Cluster Medium (ICM) detected in
X-ray observations \citep{Gunn1972ApJ, Quilis2000Sci}.  Some examples are
viscous stripping, thermal evaporation, and ram pressure stripping.  The effect
is strongest when the relative velocities are high and the ICM is very dense,
i.e. it is expected to take place in the central cluster regions. Ram pressure
stripping has been observed \citep[see][]{Cayatte1990AJ, Bravo-Alfaro2000AJ,
  Gavazzi2003ApJ} and has an indirect effect on the morphology of a galaxy as it
leads to a truncation of SF on a timescale of $\geq$ 0.1 Gyr with possibly a
small burst prior to the truncation.  When SF stops in a spiral galaxy, its disk
starts fading considerably, and the galaxy evolves towards higher bulge to disk
luminosity fractions, hence towards an earlier morphological type.  If instead
of disk gas the low density halo gas gets stripped from a galaxy falling into a
cluster gravitational potential, \emph{strangulation} has been proposed to
occur.  The halo is a gas reservoir and feeds the SF. Its removal results in a
termination of SF on a timescale of $\geq$ 1.0 Gyr. The morphology is again
influenced indirectly through the halting of SF. \emph{Galaxy mergers} are
galaxy-galaxy interactions which are most efficient at low relative velocities
\citep{Toomre1972ApJ, Mihos2004cgpc}.  Mergers are therefore more frequent in
galaxy groups than in clusters. The relative velocities of galaxies in groups
are comparable to the velocities of the stars within the galaxy, a condition
very favorable for merging, while in clusters the galaxy velocity dispersion is
higher by factors $\geq$ 10.  If the interacting galaxies are gas-rich, a major
starburst can be triggered, followed by a truncation of the star formation on a
timescale of $\sim$ 0.1 - 0.4 Gyr.  Observations about HI gas in E+A galaxies
indicate galaxy-galaxy interactions or mergers as possible transformation
mechanisms \citep{Chang2001AJ, Buyle2008arxiv}.  Enhanced merging within groups
falling into clusters has been evolved as possible explanation of the
observation that the galaxy population changes already at relatively large
distances from the cluster center around 3 R$_{vir}$
\citep[see][]{Verdugo2008A&A}, where the ICM is not yet expected to have a
strong effect.  While starbursts in the course of tidal stripping of galaxies
are at most expected to be weak, starbursts accompanying gas-rich mergers have
been found to be very strong in some cases.  All of these transformation
scenarios result in both a morphological and a spectral transformation of the
infalling spiral-rich field galaxy population.  The E+A galaxies seem to be a
missing link in the process of spiral galaxies evolving into early-type galaxies
and the investigation of this galaxy type will lead to a better understanding of
galaxy evolution. With surveys like SDSS, CNOC and LCRS, E+A galaxies were also
found in the field, which indicates that transformation of galaxies does not
only take place in clusters, but also in enhanced density environments outside
clusters \citep{Zabludoff1996, Balogh1999}.  The respective timescales for the
transformation scenarios need not to be the same and are still largely
unexplored. The Hubble Space Telescope (HST) makes it possible to also study the
morphological evolution of galaxies and compare the timescales of spectroscopic
and morphological transformation.

\subsection{E+A Galaxies}
The spectra of E+A galaxies look like a superposition of a passive continuum
spectrum of an elliptical galaxy (or an ensemble of K-type stars) without
emission lines but with deep Balmer absorption lines typical for A
stars. Therefore this class of galaxies was named E+A or k+a, which is only a
spectral but not a morphological classification.  A lack of emission lines like
in the E+A galaxies indicates a lack of actively star-forming regions
\citep{Goto2003ApJ}. Usually the equivalent width of the emission line
      [\mbox{O\,{\sc ii}}] at 3727 \AA\ is measured as an indicator of current
      star formation.  On the other hand, Balmer absorption lines are an
      indicator for recent SF. They are produced by a dominance of intermediate
      age stars in the galaxy and are strongest for A0 stars.  A-type stars
      begin to dominate the light of a galaxy about 0.5 Gyr after SF stops and
      have a lifetime of about 3 Gyr. Their signature in the spectrum can best
      be seen 1-1.5 Gyr after active SF \citep{Leonardi1996AJ,
        Poggianti2004PoS}.  This knowledge about the origin of the Balmer
      absorption lines has led to the conclusion that E+A galaxies are
      post-starforming, passive galaxies which had their last star formation
      about 1-1.5 Gyr ago \citep{Couch1987MNRAS, Poggianti2004PoS}. A starburst,
      i.e. a short-term increase in the SF activity, preceding the halting of SF
      is required in galaxies with strong Balmer lines.  The strength of the
      Balmer lines is a measure of the strength of the recent SF.

The exact definitions of E+A galaxies varies among different
authors. \citet{Yang2004} set the threshold for E+A galaxies at an EW[$\langle$
  H$\beta$, H$\gamma$, H$\delta$ $\rangle$] $>$ 5.5 \AA, while
\citet{Poggianti2004ApJ} classify galaxies with EW(H$\delta$) $>$ 3 \AA\ as E+A
galaxies and propose an additional threshold of EW(H$\delta$) $>$ 5 \AA\ for
\emph{strong} E+A galaxies.  For the EW([\mbox{O\,{\sc ii}}]), a threshold
around 2-5 \AA\ is usually adopted, depending on the detection limit.  In this
work, we chose the H$\delta$-line at 4100 \AA\ as an indicator for recent SF,
because this line is less affected by emission filling than lower order lines.
The threshold for E+A galaxies is set to EW(H$\delta$) $\geq$ 5 \AA.

\subsection{Goal of this Work}
In this work we will investigate different scenarios which are thought to be
responsible for the transformation from one galaxy type to another.  Our focus
is on E+A galaxies, which are believed to be a transition stage in galaxy
transformation from low density environment spirals to high density environment
S0 galaxies. For this purpose a grid of galaxy models is calculated with our
galaxy evolution code GALEV, which models the photometric, spectral and chemical
evolution of a galaxy.  With starbursts and/or the halting of the star formation
we can simulate the effect of the different processes of galaxy transformation
on the SFR. The GALEV code does not take into account the dynamical evolution or
spatial resolution. We only consider here the effect of the various
transformation scenarios on the SFR and on the evolution of the integral
spectrum and colors of a galaxy that it induces.

Our goal is to identify possible progenitors and successors of E+A galaxies and
investigate their spectra, luminosities and colors, in order to understand the
role of the different transformation processes.  The paper is organized as
follows. In Sect. \ref{chapter_model} we will describe our galaxy evolution code
GALEV. Sect. \ref{results} gives the results that can be drawn from our
calculated grid of undisturbed and post-starburst galaxies. In
Sect. \ref{conclusion} a conclusion is given.  In a companion paper \citep[, hereafter Paper II]{Falkenberg+09b} we will investigate the Spectral
Energy Distributions (SEDs) of our post-starburst galaxy models and make a
comparison with observations.


\section{Model Description}\label{chapter_model}
\subsection{Undisturbed Galaxies}

The GALEV code starts from a gas cloud of primordial abundance and a given
initial mass.  A given star formation rate (SFR) $\psi(t)$ determines the total
amount of stars formed in each time step.  The mass distribution of the new
stars is determined by an initial mass function (IMF) $\phi$. The evolution of
each star in the Hertzsprung-Russell diagram (HRD) is traced with a set of
stellar isochrones for 5 different metallicities.  We have used the latest
isochrones from the Padova group \citep{Bertelli1994A&AS, Girardi2003MmSAI} that
contain 26 stellar masses with 1067 evolutionary stages including the
thermal-pulsing AGB phase.  Each stage is determined by effective temperature
$T_\mathrm{eff}$, luminosity $\frac{L_\mathrm{bol}}{L_\mathrm{bol,\sun}}$ and
life time.  At any timestep the galaxy is described by a set of weighted
isochrones. With a library of stellar model atmosphere spectra, it is possible
to calculate a spectrum for each isochrone. The adopted stellar library contains
spectra from the UV to the NIR for all spectral types and luminosity classes for
5 metallicities \citep{Lejeune1998A&AS}.  Summing the isochrone spectra for each
metallicity, weighted by the SFR at the birth of the stars, and adding the
spectra of the various metallicity subpopulations, we synthesize an integrated
galaxy spectrum.  For details about the GALEV code see \citet{Anders2003A&A}.

\begin{figure}
\includegraphics[angle=270, width=84mm]{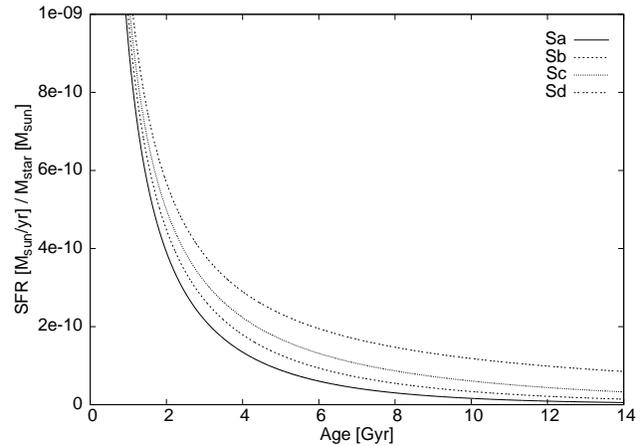}
\caption{Time evolution of the SFR of the undisturbed Sa, Sb, Sc and Sd galaxy models.}
\label{sfr_undist}
\end{figure}

For undisturbed spiral galaxy models we tie the SFR $\psi(t)$ to the evolving
gas content $G(t)$ with $G(t)$ $\propto$ $\psi(t)$ and determine efficiency
parameters for spectral galaxy types Sa, Sb, Sc, Sd as to obtain, for a Salpeter
IMF, at an age of 12 Gyr agreement with observations in the UV, optical and NIR.
The initial gas masses, and hence the initial value of the SFR are chosen as to
obtain, after 12 Gyr of evolution, the average observed B-band luminosities of
the respective galaxy types Sa, Sb, Sc, Sd as determined for Virgo by
\citet{Sandage1985AJb}.  Fig. \ref{sfr_undist} shows the evolution of the SFR
$\psi(t)$ over time for the undisturbed Sa, Sb, Sc and Sd galaxy models.

We stress that our Sa, Sb, Sc, Sd galaxy models are meant to describe
\emph{spectral galaxy types}. The observed one-to-one correspondence between
spectral and morphological galaxy types in the local universe might, of course,
not hold back until arbitrary high redshifts.

The UV, optical and NIR colors, the spectra, the gas content and the chemical
abundances are in agreement with observations as shown in \cite{Bicker2004A&A}.
However in contrast to \cite{Bicker2004A&A}, in this work the metallicity of the
stellar population is held constant at $Z=0.008$, i.e. at about half-solar
metallicity. Half-solar metallicity is a typical value for spiral galaxies in
the local and moderately distant universe \citep{Tremonti2004ApJ,
  Zaritsky1994ApJ}.  There are several reasons why closed-box models used in the
chemically consistent description do not appear appropriate for investigating
E+A galaxies: 1.) E+A are likely to be caused by major mergers of two gas-rich,
i.e. late-type spirals \citep{Kavirai2007MNRAS} in order to reproduce the large
fraction of newly formed stars. Those mergers are by definition no closed-boxes,
so that assuming closed box models are not justified. 2.) Other galaxy
transformation processes in clusters are as well expected to be accompanied by
mass loss. 3.) Bursts are likely to be associated with galactic outflows, also
making the assumptions of closed-boxes invalid. Further support comes from the
non-detections of many E+As in HI, implying low remaining gas-masses
\citep{Goto2004A&A,Liu2007ApJ}.

The GALEV code does not take into account that galaxies might contain a high
amount of dust, which has a reddening effect on the integral colors.  However,
the majority of post-starburst galaxies are believed to contain no or only
little amount of dust \citep{Balogh2005MNRAS, Goto2004A&A} and therefore can be
modelled accurately with GALEV. Some post-starburst galaxies, on the other hand
seem to still contain some dust. Those galaxies, of course, cannot appropriately
be described by our dust-free models

Our spectra do not have enough resolution to quantitatively analyze any
lines. While we implemented a calculation for the [\mbox{O\,{\sc ii}}] emission
lines into the GALEV code, for the \mbox{analysis} of the Balmer lines we used
the Lick calibration to calculate the line strength separately (see
Sect. \ref{OII_rel} and \ref{lick}).

\subsection{Lick Indices}\label{lick}
GALEV models include Lick indices, which are spectral absorption features in
well-defined wavelength ranges. The names of the 25 Lick indices refer to the
most prominent line in the wavelength range of each individual index. Most
indices are measured in terms of equivalent widths with the following
definition:

\begin{equation}
\mathrm{EW}[\AA]=\int_{\lambda_{1}}^{\lambda_{2}}\frac{F_\mathrm{C}(\lambda)-F_\mathrm{I}(\lambda)}{F_\mathrm{C}(\lambda)}d\lambda
\end{equation}

Here $F_\mathrm{I}$ is the flux for the index between two wavelengths
$\lambda_{1}$ and $\lambda_{2}$.  $F_\mathrm{C}$ is the continuum flux,
calculated from two "pseudo-continua" with wavelength ranges defined to the left
and right of the central index bandpass.

We model the Lick indices for individual stars on the basis of polynomial
fitting functions of \citet{Worthey1994ApJS} and \citet{Worthey1997ApJS}. The
index strength for the isochrones are derived by integrating and weighting with
the IMF.  With the isochrones from all contributions at each timestep, we get
weighted Lick indices for the galaxy \citep[for detail see][]{Lilly2006A&A}.

In this work, we only investigate the Balmer line indices H$\beta$, H$\gamma$,
H$\delta_\mathrm{F}$ and in particular H$\delta_\mathrm{A}$. In
Tab. \ref{tab_lick} the bandpass definitions of the index bandpasses and the
pseudo-continua are listed \citep[see][]{Trager1998ApJS}.

\begin{table}
\caption{Wavelength ranges for the Lick indices of the Balmer lines.}
\label{tab_lick}
\centering
\begin{small}
\begin{tabular}{l l l l}\hline \hline
Index & Blue continuum & Index bandpass & Red continuum \\ \hline
 H$\beta   $ & 4827.875--4847.875 & 4847.875--4876.625	& 4876.625--4891.625  \\
 H$\gamma  $ & 4283.500--4319.750 & 4319.750--4363.500	& 4367.250--4419.750  \\
 H$\delta_\mathrm{A}$ & 4041.600--4079.750 & 4083.500--4122.250	& 4128.500--4161.000  \\
 H$\delta_\mathrm{F}$ & 4057.250--4088.500 & 4091.000--4112.250	& 4114.750--4137.250  \\ \hline
\end{tabular}
\end{small}
\end{table}

\subsection{[\mbox{O\,{\sc ii}}]-Emission Lines at 3727\AA\ }\label{OII_rel}
We implemented into the code a calculation of the EW([\mbox{O\,{\sc ii}}]) for
the [\mbox{O\,{\sc ii}}]-line at 3727 \AA, which is an indicator of current SFR
and is used for the definition of E+A galaxies. The EW([\mbox{O\,{\sc ii}}]) can
be modelled from the SFR and B-band luminosity with the following relations
given by \citet{Kennicutt1992ApJ}:

\begin{equation}
\mathrm{SFR}\left(\mathrm{M_{\sun}}\ \mathrm{yr^{-1}}\right)\simeq 5 \cdot 10^{-41} L\left(\left[\mathrm{\mbox{O\,{\sc ii}}}\right]\right)
\end{equation}

and

\begin{equation}
{L\left(\left[\mathrm{\mbox{O\,{\sc ii}}}\right]\right)}\sim1.4\pm0.3\cdot10^{29}\frac{{L_\mathrm{B}}}{L_\mathrm{B}(\sun)} {\mathrm{EW}\left(\left[\mathrm{\mbox{O\,{\sc ii}}}\right]\right)}\ \mathrm{ergs^{-1}}
\end{equation}

\subsection{Galaxy Transformation Scenarios}
The interactions of galaxies, as described in Sect. \ref{intro} have different
characteristic influences on the SFR of a galaxy.  Only the effects on the SFR
will be the subject of this investigation.  The dynamical and morphological
transformation, which accompany galaxy transformation processes are beyond the
scope of this work.

The various mechanisms are realized with GALEV by inducing a sudden change in
the SFR of a normal undisturbed galaxy. Three different ways of changing the SFR
can describe the four important mechanisms of galaxy interaction. 1.) A
truncation of the SFR on a short timescale, 2.) a termination of SF on a long
timescale, and 3.) a burst, which is a rapid short-term increase in SFR before a
truncation or termination of the SFR.  Our assumption of a sudden increase is
compatible with the findings from \cite{Bekki2005MNRAS}, who used a 2-D
dynamical simulation of spiral mergers and also found a very sharp rise in SFR
at the onset of the burst followed by an exponentially declining SFR.  Possible
scenarios for the decline in SFR after a burst are exhaustion of the fuel for
SF, the heating and stirring of the ISM during the burst or the expansion of the
remaining gas by a large scale galactic wind.  We chose spiral type galaxies as
most likely progenitors for E+A galaxies. Bursts of SF at various times
$t_\mathrm{burst}$, with various burst strengths b and decline timescales $\tau$
are induced in the different spiral type models to simulate the different burst
and truncation scenarios. Tab. \ref{tab_processes} shows how the effects of the
different transformation processes are described by our models in their impact
on the SFR.

\begin{table}
\caption{Effects on the SFR and timescales for the halting of SF for galaxy interactions.}
\label{tab_processes}
\centering
\begin{tabular}{l l l }\hline \hline
Interaction   & Effect on SFR  & $\tau$ [Gyr] \\ \hline 
merger        & burst with truncation  & 0.1, 0.3         \\
harassment    & (burst and) termination& 1.0            \\
gas stripping &(burst and)  truncation & 0.1, 0.3     \\
strangulation & termination            & 1.0            \\ \hline
\end{tabular}
\end{table}

To model a burst, the SFR of an undisturbed spiral is set to a value $\psi_\mathrm{burst}$ at time $t_\mathrm{burst}$. The decline after the burst is described by an exponential law with a decline timescale $\tau$.

\begin{equation}
\psi(t\geq t_\mathrm{burst})=\frac{\psi_\mathrm{burst}}{e^{{t-t_\mathrm{burst}}/ \tau}}
\end{equation}

The maximum burst SFR $\psi_\mathrm{burst}$ together with $\tau$ determines the burst strength b. In our models b is primarily defined as the fraction of the remaining gas at $t_\mathrm{burst}$ that is consumed in the burst

\begin{equation}
b:=\frac{\Delta G}{G(t_\mathrm{burst})}
\end{equation}

$G(t_\mathrm{burst}$) is the gas mass at the beginning of the burst, $\Delta$G the fraction of $G(t_\mathrm{burst}$) that is consumed for SF during the burst.
Other definitions of burst strength also found in the literature are based on the increase of stellar mass $S$  (\ref{b_1}) or the SFR enhancement (\ref{b_2}).

\begin{equation}\label{b_1}
b_s=\frac{\Delta S}{S}
\end{equation}

or 

\begin{equation}\label{b_2}
b_{\psi}=\frac{\psi_{burst}}{\psi_i}
\end{equation}

with $\psi_i$ as the SFR just before the burst. In Tab. \ref{sfr_defs}, the burst strength values from these three definitions are compared for a few examples.

\begin{table}
\caption{Comparison of the burst strengths obtained for three different definitions in the literature.}
\label{sfr_defs}
\centering
\begin{tabular}{l l l l l l}\hline \hline
Type & $t_\mathrm{burst}$ [Gyr] &$\tau$[Gyr] & $b$ [\%] & $b_\mathrm{s}$ & $b_\mathrm{\psi}$ \\ \hline   
Sa & 11 & 0.1 & 30 & 0.01 & 7.13 \\
   &     & 1.0   & 30 & 0.01 & 0.71 \\
   &     & 1.0   & 70 & 0.02 & 1.66 \\
   & 9   & 1.0   & 30 & 0.01 & 0.60 \\
   &     & 1.0   & 70 & 0.03 & 1.40 \\
   & 6   & 1.0   & 30 & 0.04 & 0.71 \\
   &     & 1.0   & 70 & 0.10 & 1.66 \\ \hline
Sd & 11  & 1.0  & 30 & 0.24 & 2.22 \\
   &     & 1.0   & 70 & 0.56 & 5.17 \\
   & 9   & 1.0   & 30 & 0.36 & 2.71 \\
   &     & 1.0   & 70 & 0.83 & 6.33 \\
   & 6   & 1.0   & 30 & 0.68 & 3.46 \\
   &     & 1.0   & 70 & 1.58 & 8.08 \\ \hline
\end{tabular}
\end{table}

A truncation or a termination of SFR without a preceding starburst is implemented with a similar exponential law starting from the SF at the beginning of the truncation/termination $t_\mathrm{trunc}$.

We have calculated a grid for the spiral galaxy types Sa, Sb, Sc and Sd each
with burst strengths of b=0\%, 30\%, 50\% and 70\%, onsets of the bursts at
galaxy ages of 3 Gyr, 6 Gyr, 9 Gyr and 11 Gyr and decline times of 0.1 Gyr, 0.3
Gyr and 1.0 Gyr. A burst strength b=0\% stands for pure truncation or
termination of SF depending on the decline time.  In the following we will refer
to galaxies with starbursts, SF truncation and SF termination \emph{together} as
(post-)starburst galaxies, since one can describe a truncation or termination as
starburst with a burst strength of 0\%.
 
\subsection{The SFRs of Starburst Galaxy Models}
\begin{figure}
\includegraphics[width=84mm]{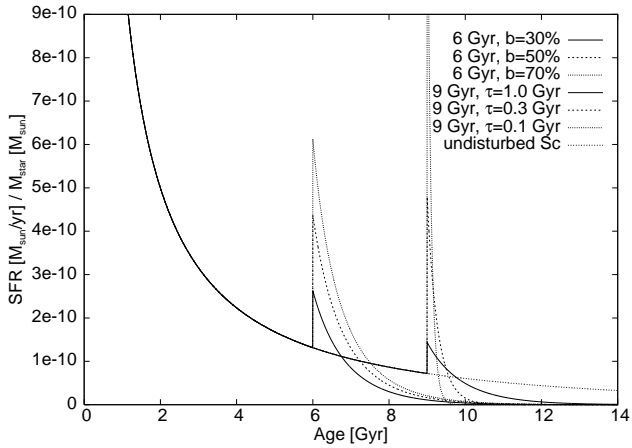}
\caption{Time evolution of the SFR of Sc galaxy models with bursts. The bursts
  beginning at 6 Gyr have different burst strengths of 30\%, 50\% and 70\%, but
  the same decline times of 1.0 Gyr. The bursts beginning at 9 Gyr have
  different decline times of 0.1, 0.3 and 1.0 Gyr, but the same burst strength
  of 30\%.}
\label{SFR_burst2}
\end{figure}

In Fig. \ref{SFR_burst2} the evolution of the SFR $\psi(t)$ for an Sc galaxy
model with various burst scenarios is shown in contrast to an undisturbed Sc
galaxy.

Three bursts beginning at 6 Gyr with the same decline times of $\tau$=1.0 Gyr
but different burst strengths of b=30\%, 50\% and 70\% are compared.  From the
three burst scenarios beginning at 9 Gyr on the right, it can be seen that at a
given gas consumption rate the decline time, i.e. the transformation scenario,
influences the maximum of the SFR in the burst significantly. At fixed $\tau$ a
burst with higher gas consumption has a larger $\psi_\mathrm{burst}$. While a
long decline time typical for mild harassment or strangulation results only in a
small maximum burst SFR $\psi_\mathrm{burst}$, a medium or short decline time
caused by gas stripping or by a merger produces much higher burst SFRs when
consuming the same amount of gas.  It can also be seen that the burst SFR
significantly depends on the onset time of the burst. A later beginning implies
that the galaxy has already consumed more of its gas content. Therefore a burst
at a galaxy age of 9 or 11 Gyr will not have as high a SFR as a burst at 6 Gyr.

\begin{figure}
\includegraphics[width=84mm]{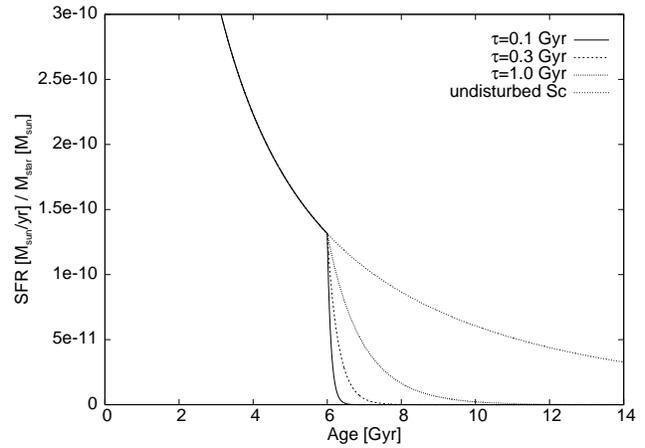}
\caption{Time evolution of the SFR of Sc type spiral galaxies with a halting of
  SFR on different timescales of 0.1, 0.3 and 1.0 Gyr, but with the same onset
  at 6 Gyr.}
\label{SFR_burst3}
 
\end{figure}

The scenarios in Fig. \ref{SFR_burst3}, again compared to an undisturbed Sc
model, show the truncation/termination of the SF of an Sc model at an age of 6
Gyr on different timescales $\tau$. The models with $\tau$=0.1 Gyr and
$\tau$=0.3 Gyr simulate a truncation after a merger or the stripping of gas from
the disk.  The model with $\tau$=1.0 Gyr is a termination of SF as caused by
galaxy harassment or strangulation.

\section{Results}\label{results}

\subsection{Lick Index H$\delta$ of Undisturbed Galaxies}\label{chapter_Hd}
\begin{figure}
\includegraphics[width=84mm]{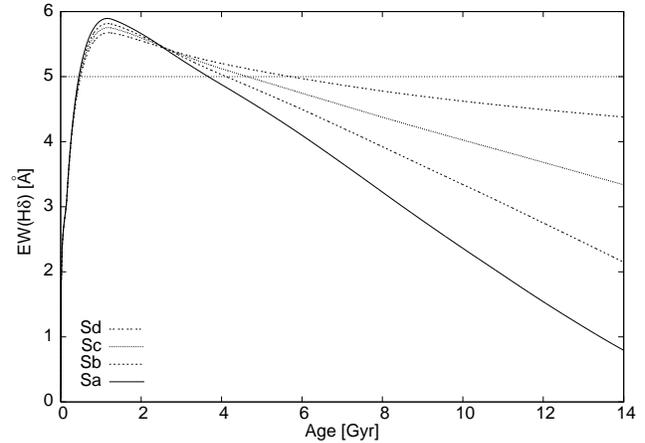}
\caption{Time evolution of the EW(H$\delta$) of undisturbed spiral galaxies. The horizontal line marks the 5 \AA\ threshold.}
\label{Hd_undist}
 
\end{figure}

In this Sect. we investigate the evolution of the EW(H$\delta$) for our
undisturbed galaxy models for spectral types Sa, Sb, Sc to later compare them
with the transformation models.  In Fig. \ref{Hd_undist} we plot the
EW(H$\delta$) as a function of galaxy age.  The horizontal line in
Fig. \ref{Hd_undist} indicates the 5 \AA\ threshold, separating H$\delta$-strong
from non-H$\delta$-strong galaxy models. Galaxies lying above this line during
any time of their evolution would be observationally classified as
H$\delta$-strong galaxies and therefore fulfill one of the two criteria for the
definition of E+A galaxies.

Fig. \ref{Hd_undist} shows that the EW(H$\delta$) grows fast after the onset of
SF and reaches a maximum value around 1 Gyr.  The EW(H$\delta$) of 5.5--6
\AA\ of the different galaxy types declines on different timescales.

It is to note that all undisturbed normal spiral galaxy models lie above the
threshold and therefore have an H$\delta$-strong phase early in their evolution
lasting until ages of 3 to almost 6 Gyr, i.e. to redshifts $z\sim$2--1 for Sa
and Sd types, respectively.  The values for EW([\mbox{O\,{\sc ii}}]) and for
EW(H$\delta$) for an Sa and an Sd model at 1.5 Gyr are listed in
Tab. \ref{EW_tab}. The values for the EW([\mbox{O\,{\sc ii}}]) are too high for
the undisturbed galaxies in the H$\delta$-strong phase to classify them as E+A
galaxies.

\begin{table}
\caption{EW([\mbox{O\,{\sc ii}}]) and EW(H$\delta$) of Sa and Sd model at ages of 1.5 Gyr, 4 Gyr (Sa) and 6 Gyr (Sd), respectively.}\label{EW_tab}
\centering
\begin{tabular}{l l l l}\hline \hline
Type & Age [Gyr]& EW([\mbox{O\,{\sc ii}}])[\AA] & EW(H$\delta$)[\AA]\\ \hline 
Sa       & 1.5    & 37.25     & 5.83\\
	 & 4      & 26.98     & 4.88\\
Sd 	 & 1.5    & 42.06     & 5.64\\
	 & 6	  & 33.63     & 4.97\\ \hline
\end{tabular}
\end{table}

The values for EW(H$\delta$) and EW([\mbox{O\,{\sc ii}}]) between ages of 1.5
and 6 Gyr are a striking result, because it indicates that galaxies which only
follow the normal evolution, all go through a phase in which they have strong
H$\delta$-lines without having a starburst. Not taking this result into account
in analyses of high redshift galaxies leads to an underestimate of their
emission lines, which are influenced by the Balmer absorption lines, for example
the $H{\alpha}$ and $H{\beta}$, and of the SFR derived from those.

From Fig. \ref{Hd_undist} it is also obvious that a lower threshold of 3
\AA\ for EW(H$\delta$) as sometimes used in literature is not appropriate
because the undisturbed galaxy models have values of EW(H$\delta$) $\geq 3$
\AA\ for at least 9 Gyr, i.e. until $z\sim$0.4.

\subsection{Lick Index H$\delta$ for Post-Starburst Galaxies}\label{Hd}

For the discussion of the E+A galaxies, we start with the investigation of
EW(H$\delta$) and EW([\mbox{O\,{\sc ii}}]) of our starburst models, because the
presence of strong Balmer absorption lines and the absence of emission lines are
the two criteria to distinguish between E+A and non-E+A galaxy models.  In
Sect. \ref{Hd} we investigate which of our models are H$\delta$-strong by
looking at the EW(H$\delta$) as a function of time and in
Sect. \ref{chapter_[OII]} we investigate the time evolution of EW([\mbox{O\,{\sc
      ii}}]).

\begin{figure}
\includegraphics[width=84mm]{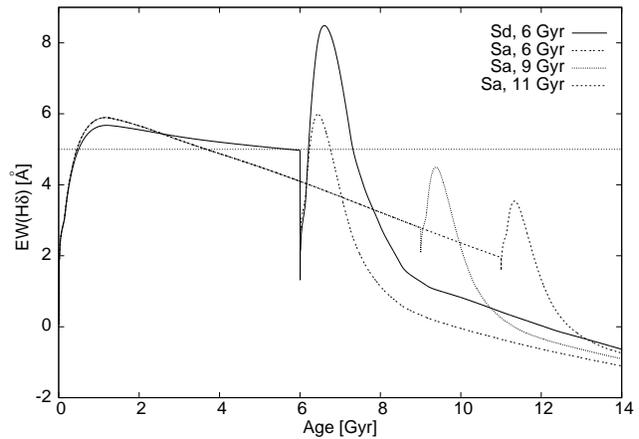}
\caption{Time evolution of EW(H$\delta$) in galaxies with different burst
  scenarios (Sa, burst of 70\% at 6 Gyr, decline time 0.1 Gyr; Sd, burst of 70\%
  at 6 Gyr, decline time 0.1 Gyr; Sa, burst of 70\% at 11 Gyr, decline time 0.1
  Gyr).}
\label{Hd_1_paper}
\end{figure}

\begin{figure}
\includegraphics[width=84mm]{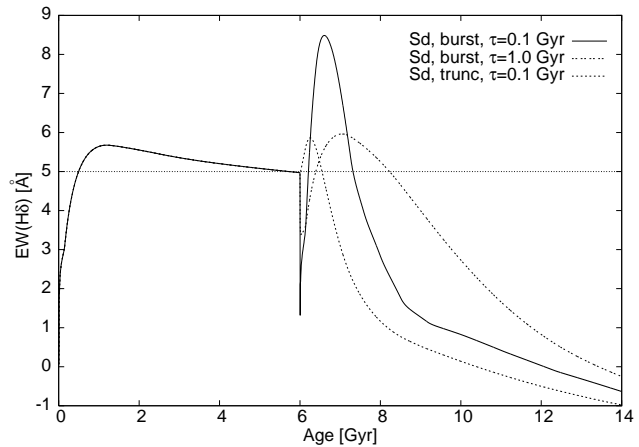}
\caption{Time evolution of EW(H$\delta$) in galaxies with different burst
  scenarios (Sd, burst of 70\% at 6 Gyr, decline time 0.1 Gyr; Sd, burst of 70\%
  at 6 Gyr, decline time 1.0 Gyr; Sd, truncation at 6 Gyr, decline time 0.1
  Gyr).}
\label{Hd_2_paper}
\end{figure}

In Figs. \ref{Hd_1_paper} and \ref{Hd_2_paper} the evolution of the
EW(H$\delta$) of different burst and truncation models are represented.
Fig. \ref{Hd_1_paper} shows an Sd and an Sa model with a burst at a galaxy age
of 6 Gyr as well as an Sa model with a burst at a galaxy age of 9 Gyr and 11
Gyr. For all four models the burst strength b amounts to 70\% and the decline
time is $\tau$=0.1 Gyr.  For the two models with a burst at 6 Gyr, the
EW(H$\delta$) drops strongly at the onset of the burst before it grows and
crosses the threshold of 5 \AA. This drop is caused by the dominance of young
burst stars, which do not show strong Balmer lines. The H$\delta$-strong phase
starts 0.2 Gyr after the onset of the burst and lasts for 0.5 Gyr (Sa) to 1.1
Gyr (Sd).  After the H$\delta$-strong phase the EW(H$\delta$) declines very fast
and becomes smaller than in the undisturbed galaxies. In contrast to the
undisturbed galaxies, the decline of the EW(H$\delta$) is similar for the
different galaxy models.  The Sd galaxy shows the strongest H$\delta$-line,
because an Sd galaxy at an age of 6 Gyr has a higher gas content than the other
spiral models which consume their gas faster and hence allows for the highest
burst SFR.  Also at an age of 6 Gyr the underlying stellar population before the
burst has already a higher EW(H$\delta$) than earlier spiral models.

Comparing the three Sa models with the different onset times of 6 Gyr, 9 Gyr and
11 Gyr, Fig. \ref{Hd_1_paper} shows that the burst strength reflected in the
strength of the H$\delta$-lines depends strongly on the gas content and its
evolution over time. For early bursts, the gas reservoir in the Sa model is
still high enough to allow for high burst SFRs and to drive the galaxy to high
values of EW(H$\delta$) after the burst.  At a galaxy age of $\leq$9 Gyr, the
gas content of the Sa galaxies is very small because most of the gas has already
been converted into stars. Due to the small amount of gas left in the galaxies,
only a small starburst can occur at these redshifts and the models do not reach
the H$\delta$-strong regime any more at low redshifts. This helps us to
understand why relatively few E+A galaxies are observed at low redshifts.

Fig. \ref{Hd_2_paper} compares burst and truncation scenarios and bursts with
different decline timescales.  In comparison to the burst scenarios, the
EW(H$\delta$) of the truncation and termination scenarios increases immediately,
without dropping first. However, the strength of EW(H$\delta$) is much weaker
than for galaxies with a preceding burst. In fact only the very strongest cases
of the truncation scenario, i.e. the sudden halting of the relatively high SFR
in Sc and Sd models, result in EW(H$\delta$) values slightly above 5 \AA\ for
0.3 Gyr (Sc) and 0.5 Gyr (Sd), while no termination scenario drives the
EW(H$\delta$) values above the threshold.  Compared to the burst scenarios, the
H$\delta$-strong phase of the truncation scenarios starts earlier, e.g. 20 Myr
after the onset of SF truncation in an Sd model.  Fig. \ref{Hd_2_paper} also
illustrates on the example of an Sd model that a longer decline time results 1.)
in weaker Balmer lines and 2.) possibly in a longer H$\delta$-strong phase if
enough gas is available and therefore the burst SFR can be strong enough for the
model galaxy to reach the 5 \AA\ threshold in the post-starburst phase.

The H$\delta$-strong phase can last up to 2 Gyr in cases with long burst timescales.

The EW(H$\delta$) strength does not simply depend on the burst strength, but on
a complex interplay of burst strength, decline timescale, onset of the burst and
progenitor galaxy type. Depending on the individual model either one of these
factors can dominate the influence on the EW(H$\delta$).

\begin{figure} 
\includegraphics[width=84mm]{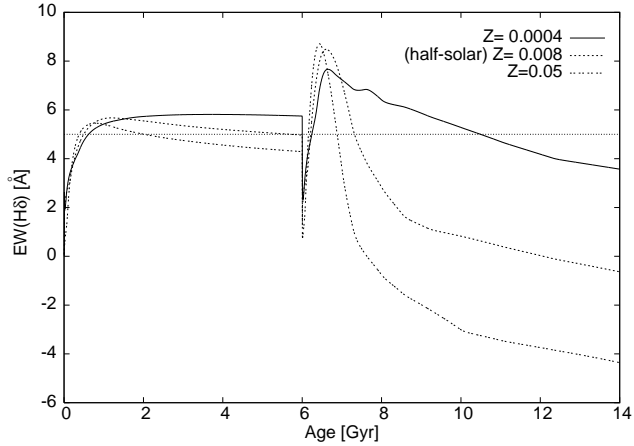} 
\caption{Evolution of H$\delta$ with time for Sd models with three different metallicities (Sd, 70\% bursts at 6 Gyr, decline times 0.1 Gyr).}
\label{Z_fig}
\end{figure}

To study the possible impact of metallicity effects on the H$\delta$-strength,
we run Sd type constant SFR models with different metallicities $0.0004\leq Z
\leq 0.05$. The scenarios in Fig. \ref{Z_fig} show three bursts at 6 Gyr, a
burst strength of b=70\%, a decline time $\tau$=0.1 Gyr, but with different
metallicities.

It is obvious from Fig. \ref{Z_fig} that the EW(H$\delta$) depends on the
metallicity of the galaxy. While the peak values achieved for EW(H$\delta$) are
somewhat higher for high metallicity galaxies (EW(H$\delta$) $\sim$ 9 \AA\ for
Z=0.05 versus EW(H$\delta$) $\sim$ 8 \AA\ for Z=0.0004), \emph{a low metallicity
  galaxy remains much longer in the H$\delta$-strong phase}.  At low
metallicities the stars are generally hotter and brighter. Therefore stars of
lower mass populate regions in the HRD that are described by A and early F
spectra: For example at solar metallicity the 1.4 $M_{\sun}$ star is the lowest
mass reaching into the early F-star range. At low metallicity, longer-living
stars down to 0.9 $M_{\sun}$ reach this range and live there for about 4
Gyr. Stars with 0.9 $M_{\sun}$ are much more numerous than 1.4 $M_{\sun}$ stars
as a consequence of the stellar IMF.  The A-star temperature range in the HRD is
reached at solar metallicities by stars $\geq$ 1.6 $M_{\sun}$ which live there
for $<$ 1.5 Gyr. At low metallicities (Z$<$0.02) these temperatures are already
reached by stars $\geq$ 1 $M_{\sun}$ which live there for $>$ 3.8 Gyr. The life
time of an A star increases therefore by a factor $\geq$ 2 at lower
metallicities, while additionally there is a higher number of A stars, both
acting together to cause a longer H$\delta$-strong phase.  These two effects
apparently strongly overcompensate the fact that the lifetime of 1.5--3
$M_{\sun}$ stars is longer at higher metallicities.  The lifetime of stars in
this mass range at a metallicity of Z=0.05 is only a factor $\sim$ 2 higher than
their lifetime at Z=0.0004 and these high mass stars are much less numerous.

Because galaxies in the early Universe tend to have lower metallicities than
galaxies in the local Universe and since a lower metallicity implies a longer
H$\delta$-strong phase, the probability to observe a galaxy in its E+A phase
increases towards higher redshifts. Therefore a higher number of E+A galaxies
cannot be ascribed solely to a higher number of starburst galaxies but also to
the lower metallicities and therefore a longer H$\delta$-strong phase of
galaxies in the early Universe. Higher maximal burst strengths enabled by the
higher gas content add to this effect.

In the GALEV code, two different Lick indices for H$\delta$ are implemented,
H$\delta_A$ and H$\delta_F$. The index H$\delta_A$, which is mainly used in this
work has a wider index bandpass and wider pseudo-continuum ranges
\citep{Lilly2006A&A}. The values of EW(H$\delta_F$) are significantly lower at
all times than the values of EW(H$\delta_A$). Therefore it is very important to
check for the consistency of the EW(H$\delta$) definitions before comparing with
observations of all the Balmer lines EW(H$\beta$), EW(H$\delta$) and
EW(H$\gamma$). The EW(H$\delta$) shows the highest values and is therefore
easier to measure than the other Balmer lines. Moreover, in contrast to lower
order Balmer lines the H$\delta$-line is less affected by any potential emission
line contributions in case of low level residual SF in the galaxy.

\subsection{H$\delta$-Strong Galaxies}\label{Hd_strong}
In this Sect. we will explore those scenarios from our grid which become
H$\delta$-strong and identify the progenitors of the H$\delta$-strong galaxies.

As shown in Sect. \ref{chapter_Hd}, all burst scenarios with a burst at 3 Gyr
are H$\delta$-strong since already the undisturbed galaxy models lie above the 5
\AA\ threshold at this time of their evolution (see Fig. \ref{Hd_undist}).
\emph{For a certain type of progenitor spiral galaxy, the essential parameter to
  cause an EW(H$\delta$) above 5 \AA\ is the peak burst SFR
  $\psi_\mathrm{burst}$ and not the burst strength. For different spiral
  progenitors, different peak burst SFRs are required in order to make them
  H$\delta$-strong after the burst.}  For example, all Sa galaxies with a burst
SFR $\psi_\mathrm{burst} \gtrsim 15 \frac{M_{\sun}}{\mathrm{yr}}$ become
H$\delta$-strong, while for Sa models with a burst SFR below this value, the
EW(H$\delta$) does not reach the 5 \AA\ threshold.  There are a few exceptions
of galaxies which have a value for $\psi_\mathrm{burst}$ above the threshold
value for the corresponding galaxy type but do not become H$\delta$-strong.
These exceptions are due to the EW(H$\delta$) of the underlying undisturbed
galaxy models. Bursts at a later stage in the life of the galaxy cannot cause as
strong an H$\delta$-line, not even for higher SFRs than bursts at earlier times,
because the EW(H$\delta$) of the underlying undisturbed model is already
low. The later the bursts start the lower are the peak values for the
EW(H$\delta$).  The EW(H$\delta$) of the underlying galaxy is also responsible
for the fact that the different galaxy types have to be examined separately and
no global minimal value can be given for a burst SFR to cause an
H$\delta$-strong phase.

Without exception all Sd models with a burst become H$\delta$-strong. The
maximum value for EW(H$\delta$) of 8.5 \AA\ in our sample is reached by the Sd
model with a burst of strength b=70\% at 6 Gyr and a decline time $\tau$=0.1
Gyr.  Almost all Sc burst models become H$\delta$-strong. Only Sc burst models
with long decline time which have moderate to low burst strength and the burst
at late galaxy ages do not reach the threshold.  Sb burst models with a long
decline time do not become H$\delta$-strong, while all Sb models with a short
decline time have a H$\delta$-strong phase. For Sb burst models with a moderate
decline time of 0.3 Gyr it delicately depends on the beginning and the strength
of the burst.  Most Sa burst models do not become H$\delta$-strong. Only very
few models with early bursts which have short decline times have a
H$\delta$-strong phase.

For the models with SF truncation and termination without a preceding burst only
Sc and Sd galaxies can possibly become H$\delta$-strong. No Sa or Sb truncation
or termination model at any galaxy age or with any decline time reaches
EW(H$\delta$)$\geq5$ \AA.  Sc truncation and termination models without a
preceding burst only become H$\delta$-strong, when SF is truncated at early ages
and on short timescales, i.e. harassment and strangulation scenarios without a
burst can be excluded as possible progenitors for E+A galaxies.  In contrast,
the Sd models with short and moderate decline times become
H$\delta$-strong. Even an Sd model with a SF termination on a long timescale at
early ages has an H$\delta$-strong phase.  Tab. \ref{z_0.4} shows a compilation
of the H$\delta$-strong models with starbursts at 9 Gyr.

\begin{table}
\caption{Compilation of H$\delta$-strong galaxy models with a burst beginning at 9 Gyr.}
\label{z_0.4}
\centering
\begin{tabular}{l l l }\hline \hline
Type & b[\%]  & $\tau$ [Gyr]  \\ \hline 
Sa   & -  &  -            \\ \hline
Sb   & 30 & 0.1           \\ 
     & 50 & 0.1           \\ 
     & 70 & 0.1, 0.3      \\ \hline
Sc   & 30 & 0.1, 0.3      \\
     & 50 & 0.1, 0.3      \\
     & 70 & 0.1, 0.3, 1.0 \\ \hline
Sd   & 0  & 0.1, 0.3      \\ 
     & 30 & 0.1, 0.3, 1.0  \\
     & 50 & 0.1, 0.3, 1.0  \\
     & 70 & 0.1, 0.3, 1.0 \\ \hline
\end{tabular}
\end{table}

It can be seen that the later the progenitor galaxy type, the more models become H$\delta$-strong.

\subsection{[\mbox{O\,{\sc ii}}]-Lines of H$\delta$-Strong Galaxies}\label{chapter_[OII]}

\begin{figure}
\includegraphics[width=84mm]{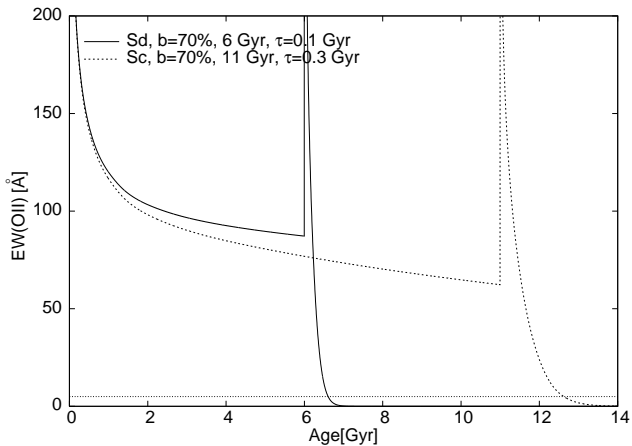}
\caption{Time evolution of EW([\mbox{O\,{\sc ii}}]) of an Sd galaxy model (Sd,
  burst 70\% at 6 Gyr, decline time 0.1 Gyr) and an Sc galaxy model (Sc, burst
  70\% at 11 Gyr, decline time 0.3 Gyr).}
\label{[OII]}
\end{figure}

The second important criterion for a galaxy to be classified as E+A galaxy is to
ensure that the galaxy has no current SF by measuring the EW([\mbox{O\,{\sc
      ii}}]) at 3727 \AA.  For this purpose we modelled the EW([\mbox{O\,{\sc
      ii}}]) on the basis of the SFR and B-band luminosity using a relation
given by \citet{Kennicutt1992ApJ} as introduced in Sect. \ref{OII_rel}.  In
Fig. \ref{[OII]}, the time evolution of the EW([\mbox{O\,{\sc ii}}]) of an Sd
galaxy with a burst of 70\%, an onset at 6 Gyr and a decline time of 0.1 Gyr,
and of an Sc galaxy with a burst of 70\%, an onset at 11 Gyr and a decline time
of 0.3 Gyr is shown. The EW([\mbox{O\,{\sc ii}}]) increases instantly at the
onset of the burst and afterwards decreases rapidly.

\begin{figure}
 
\includegraphics[width=84mm]{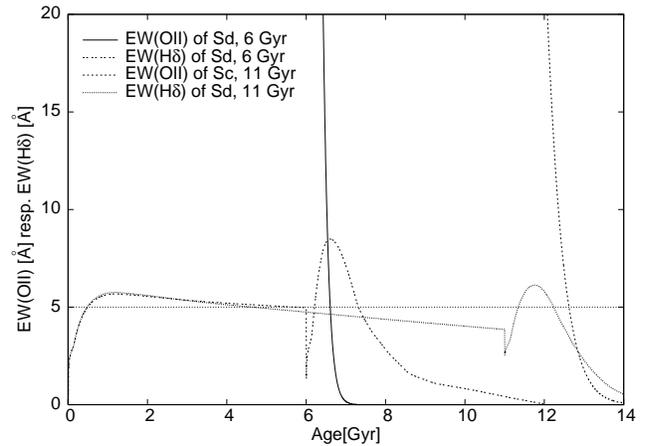}
\caption{Time evolution of EW([\mbox{O\,{\sc ii}}]) and EW(H$\delta$) of an Sd
  galaxy model (Sd, burst 70\% at 6 Gyr, decline time 0.1 Gyr) and an Sc galaxy
  model (Sc, burst 70\% at 6 Gyr with decline time 0.3 Gyr).}
\label{OII_Hd_paper}
 
\end{figure}

It is obvious that the EW([\mbox{O\,{\sc ii}}]) reflects the time evolution of
the SFR. It is to note, however, that the EW([\mbox{O\,{\sc ii}}]) remains at
high values for some time after the burst and therefore extends into parts of
the H$\delta$-strong phase, as can be seen in Fig. \ref{OII_Hd_paper} for the Sd
and Sc galaxy models described above.  In this Fig., the time evolution of
EW([\mbox{O\,{\sc ii}}]) is plotted together with the time evolution of
EW(H$\delta$). The peaks of the EW([\mbox{O\,{\sc ii}}]) are not shown in this
figure, because the values are much higher than EW(H$\delta$) during the burst
and would make it impossible to see the EW(H$\delta$) in the same plot.  The
horizontal line marks the thresholds for EW(H$\delta$) as well as for
EW([\mbox{O\,{\sc ii}}]).  The threshold for EW([\mbox{O\,{\sc ii}}]) is chosen
to be 5 \AA, i.e at the upper limit for this threshold from
\citet{Dressler1999}. We chose this upper limit taking into account that
observed spectra always have noise, which tends to bias observed
EW([\mbox{O\,{\sc ii}}]) towards lower values, while we calculated the
EW([\mbox{O\,{\sc ii}}]) from the SFR.

The Sd galaxy in Fig. \ref{OII_Hd_paper} is the model with the longest E+A phase
of 0.71 Gyr in our grid. It is the same galaxy that has the strongest peak value
for EW(H$\delta$) in our sample. The beginning of the E+A phase is given by the
drop of EW([\mbox{O\,{\sc ii}}]) below 5 \AA, while the end of the E+A phase is
defined by the drop of EW(H$\delta$) below 5 \AA. The duration of the E+A phase
thus amounts only 64\% of the H$\delta$-strong phase with a duration of 1.11
Gyr. This example shows that the E+A phase only accounts a dramatically short
fraction of the post-starburst phase.  For many galaxy models, the phase of
strong [\mbox{O\,{\sc ii}}] emission is even longer than the H$\delta$-strong
phase as demonstrated in the left of Fig. \ref{OII_Hd_paper}. The
H$\delta$-strong phase of the Sc galaxy model is already over when the
EW([\mbox{O\,{\sc ii}}]) drops below 5 \AA\ and this model therefore would not
be classified as an E+A galaxy.

It is striking that the models can clearly be separated according to the
different decline times from looking at the strength of EW([\mbox{O\,{\sc
      ii}}]).  \emph{All H$\delta$-strong galaxy models with SFR decline times
  of 0.1 Gyr have an E+A phase, i.e. sufficient low [\mbox{O\,{\sc ii}}]-lines
  during their H$\delta$-strong phase, while all models with SFR decline times
  of 1.0 Gyr and 0.3 Gyr never go through an E+A phase since they still have
  [\mbox{O\,{\sc ii}}] emission during their H$\delta$-strong phase.}

Therefore we state that the conventional definition of E+A galaxies based on the
absence of [\mbox{O\,{\sc ii}}] emission above some threshold is too narrow to
encompass the full range of post-starburst galaxies. In particular, galaxies
with a long decline time are excluded since they exhibit significant
[\mbox{O\,{\sc ii}}] emission in addition to their strong Balmer absorption
lines.  Possible processes leading to E+A galaxies can only be galaxy mergers
and gas stripping (with a starburst), while harassment and strangulation can be
excluded due to their long decline timescales.

\subsection{Spectral evolution of E+A Galaxies}\label{chapter_specs}

\begin{figure}
\includegraphics[width=84mm]{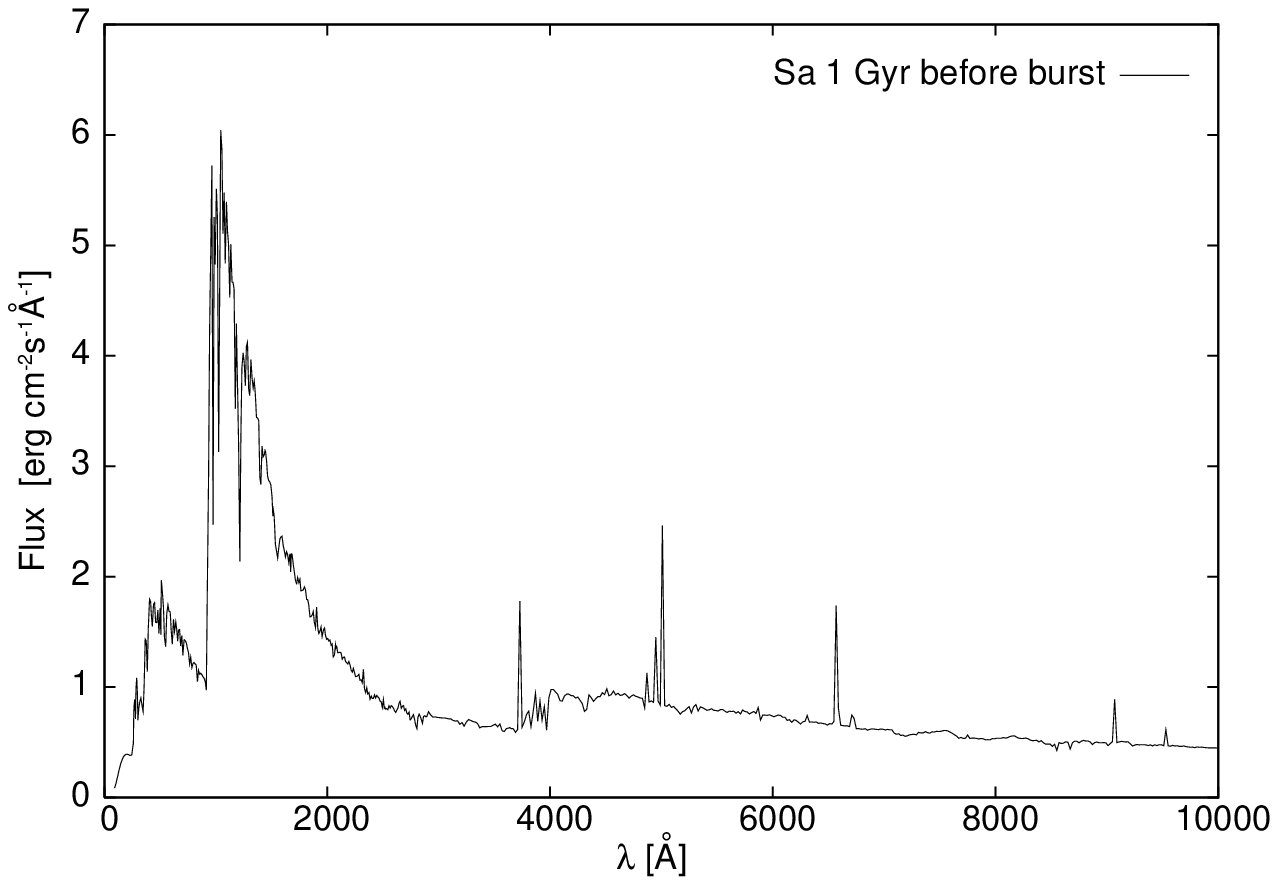}
\includegraphics[width=84mm]{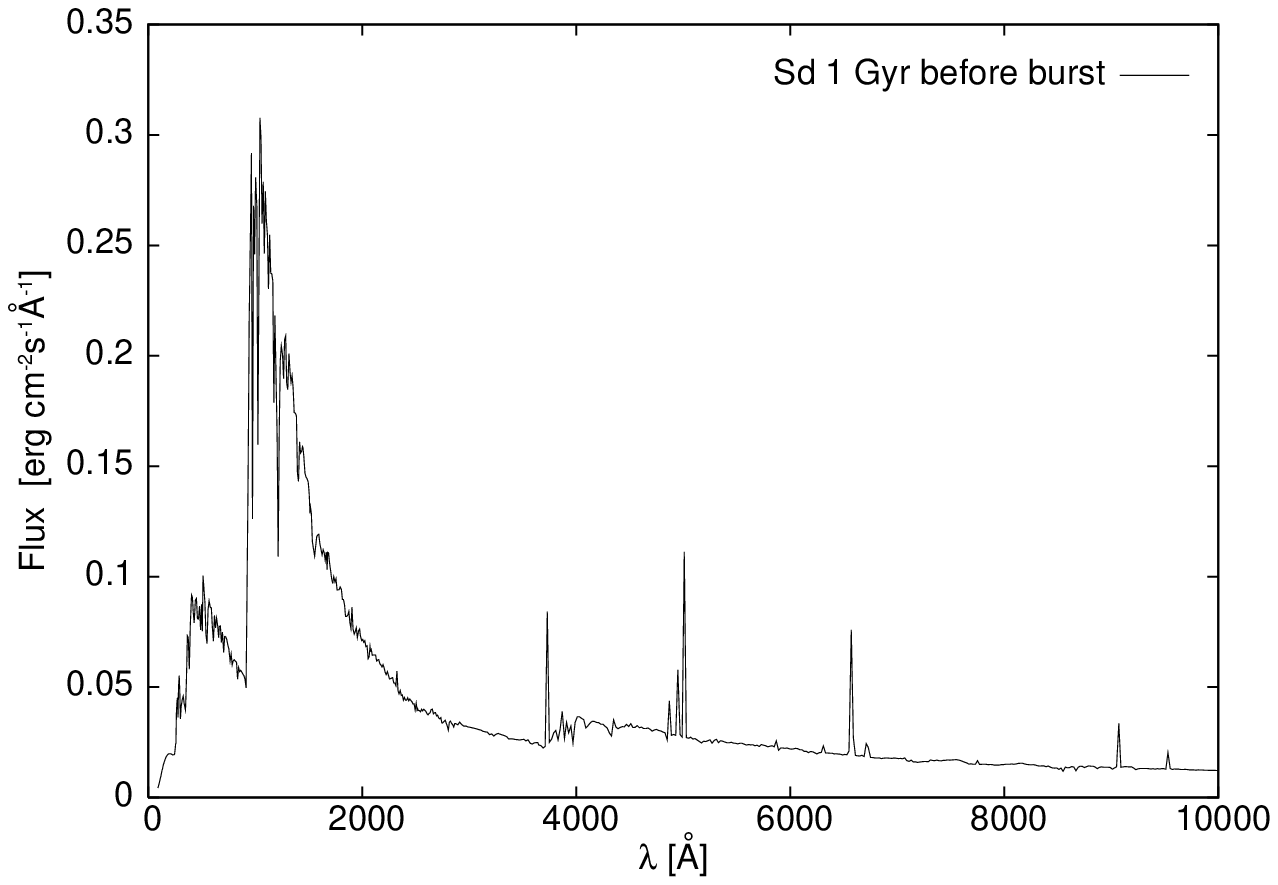}
\caption{Spectra of an Sa model (burst 50\%) and an Sd model (burst 70\%) 1 Gyr before the bursts (at 6 Gyr, decline times 0.1 Gyr).}
\label{burst_spectra_1}
\end{figure}

\begin{figure}
\includegraphics[width=84mm]{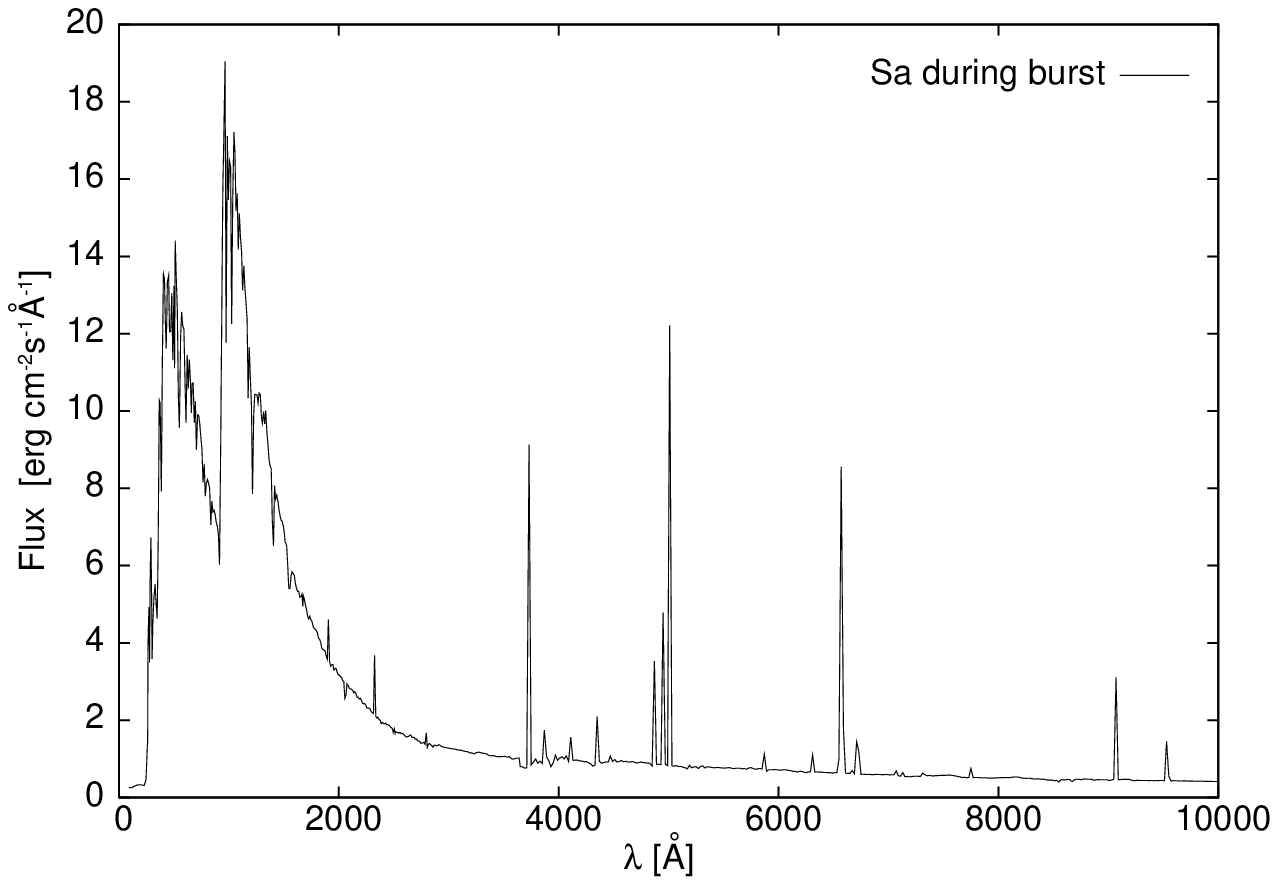}
\includegraphics[width=84mm]{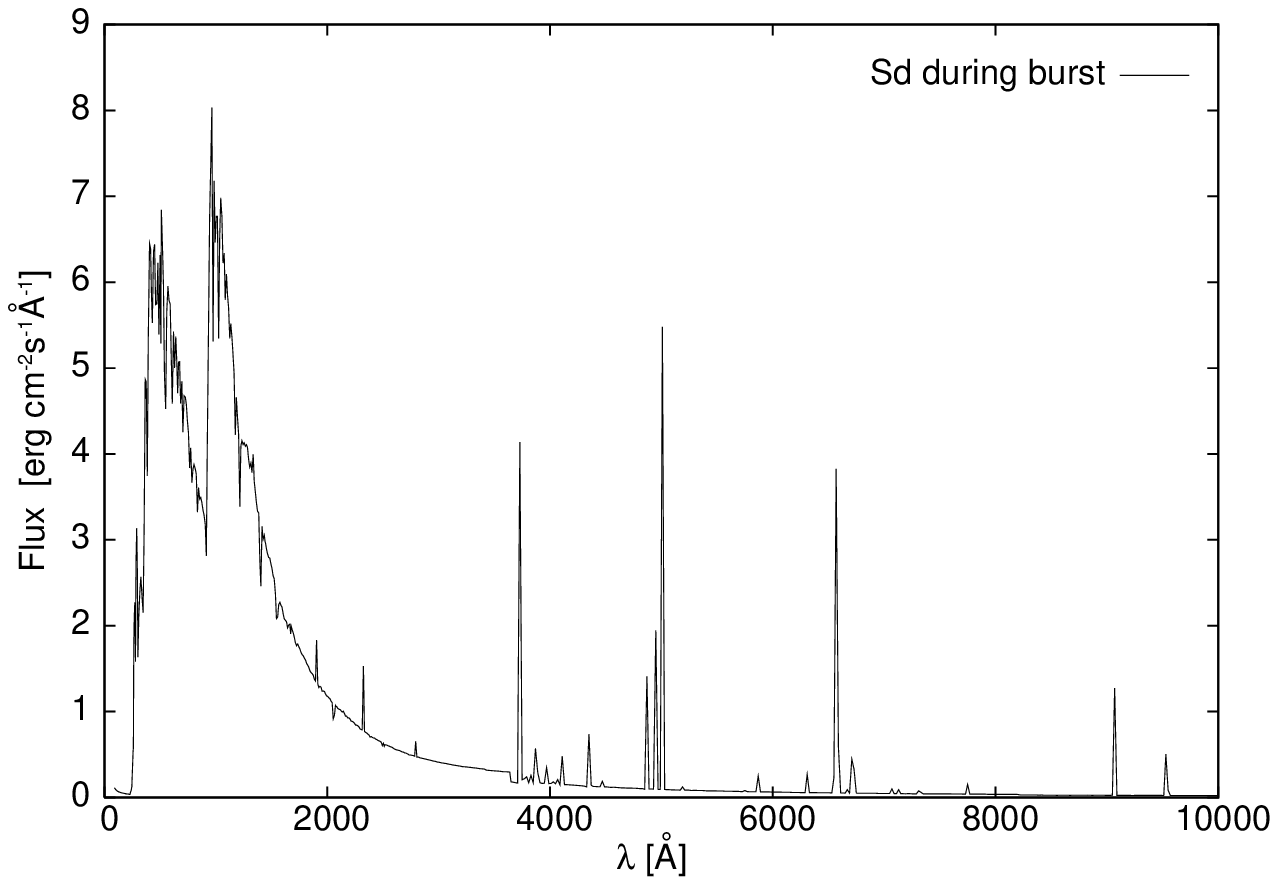}
\caption{Spectra of an Sa model (burst 50\%) and an Sd model (burst 70\%) during the burst (at 6 Gyr, decline times 0.1 Gyr).}
\label{burst_spectra_2}
\end{figure}

\begin{figure}
\includegraphics[width=84mm]{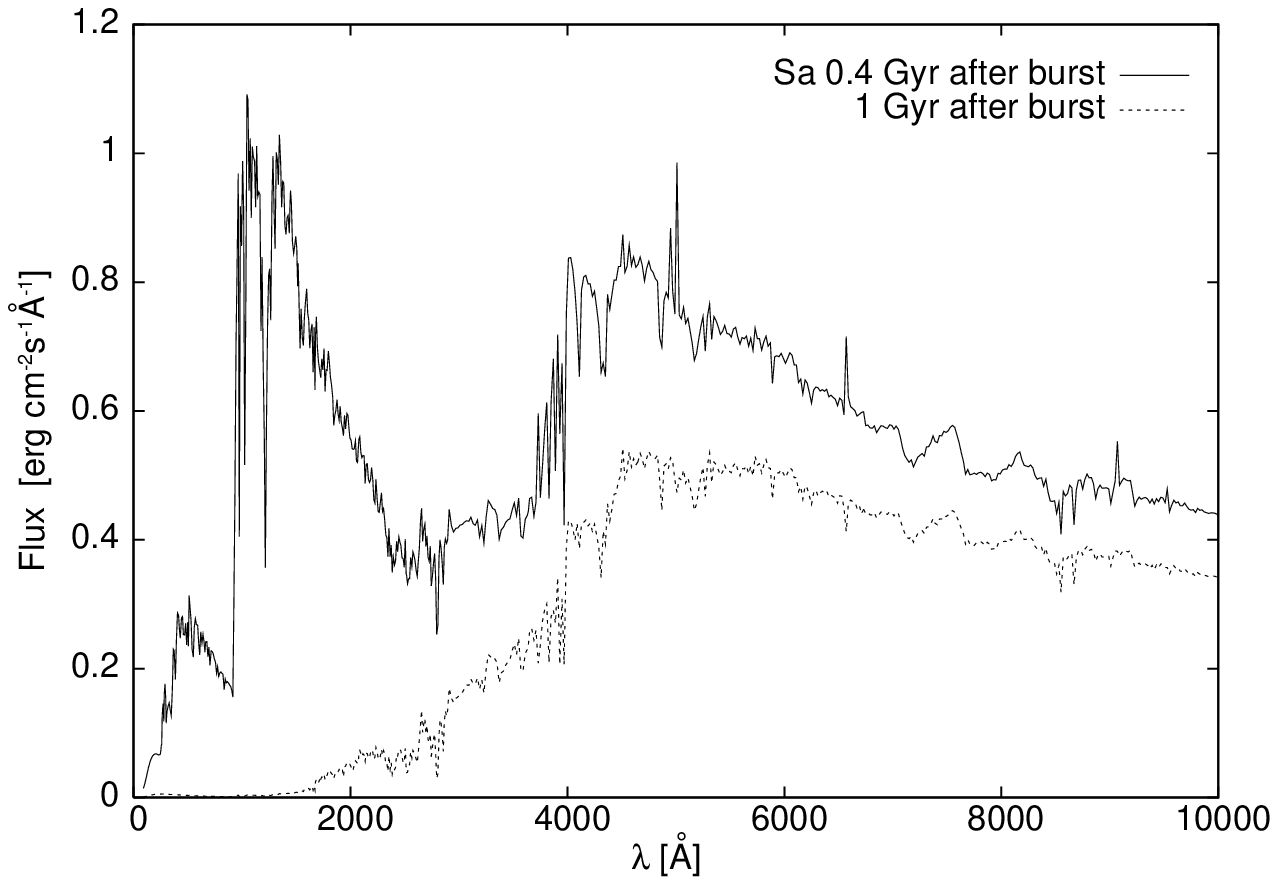}
\includegraphics[width=84mm]{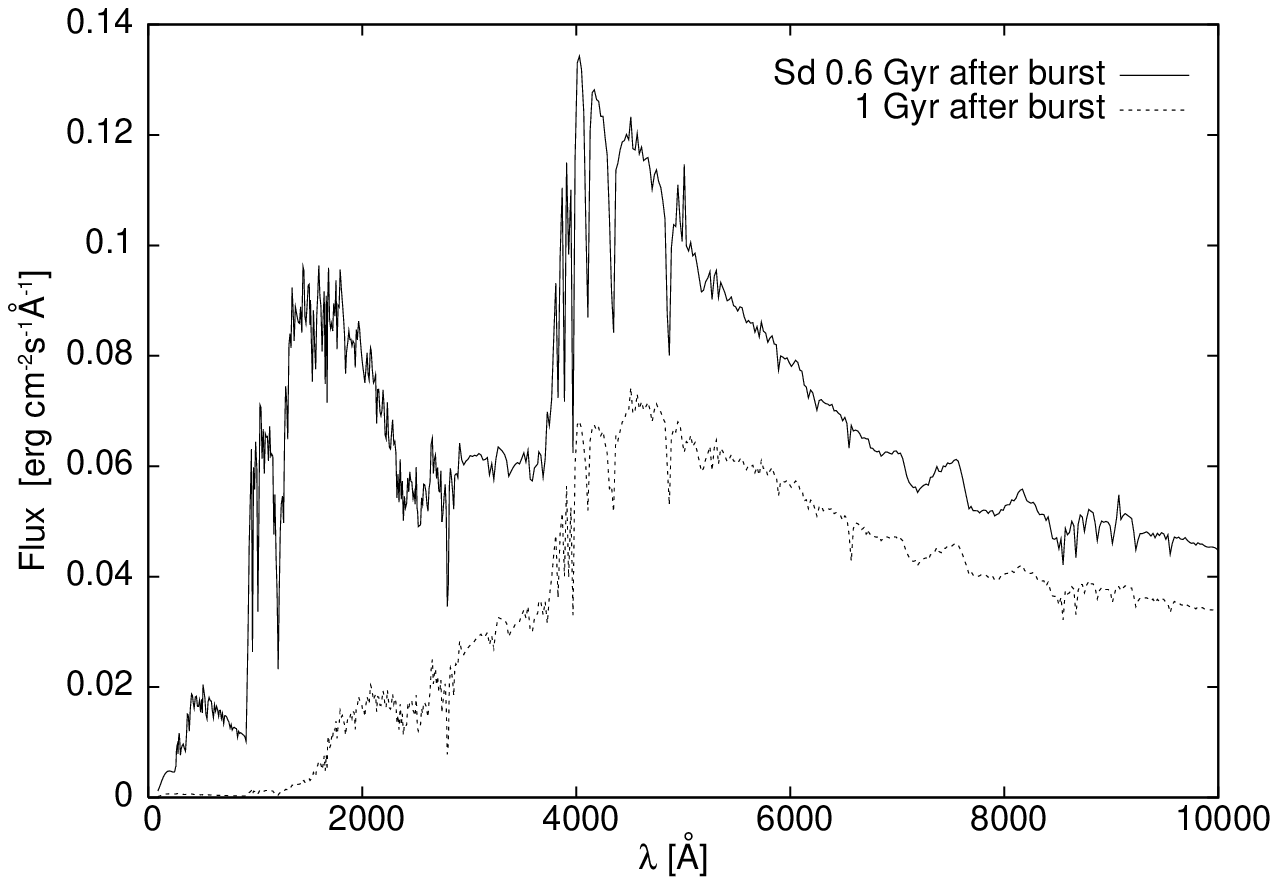}
\caption{Spectra of an Sa model (burst 50\%) and an Sd model (burst 70\%) in the burst (at 6 Gyr, decline times 0.1 Gyr). The spectra show the galaxies in their post-starburst phase at their maximum EW(H$\delta$) and 1 Gyr after the burst.}
\label{burst_spectra_3}
\end{figure}

\begin{figure}
\includegraphics[width=84mm]{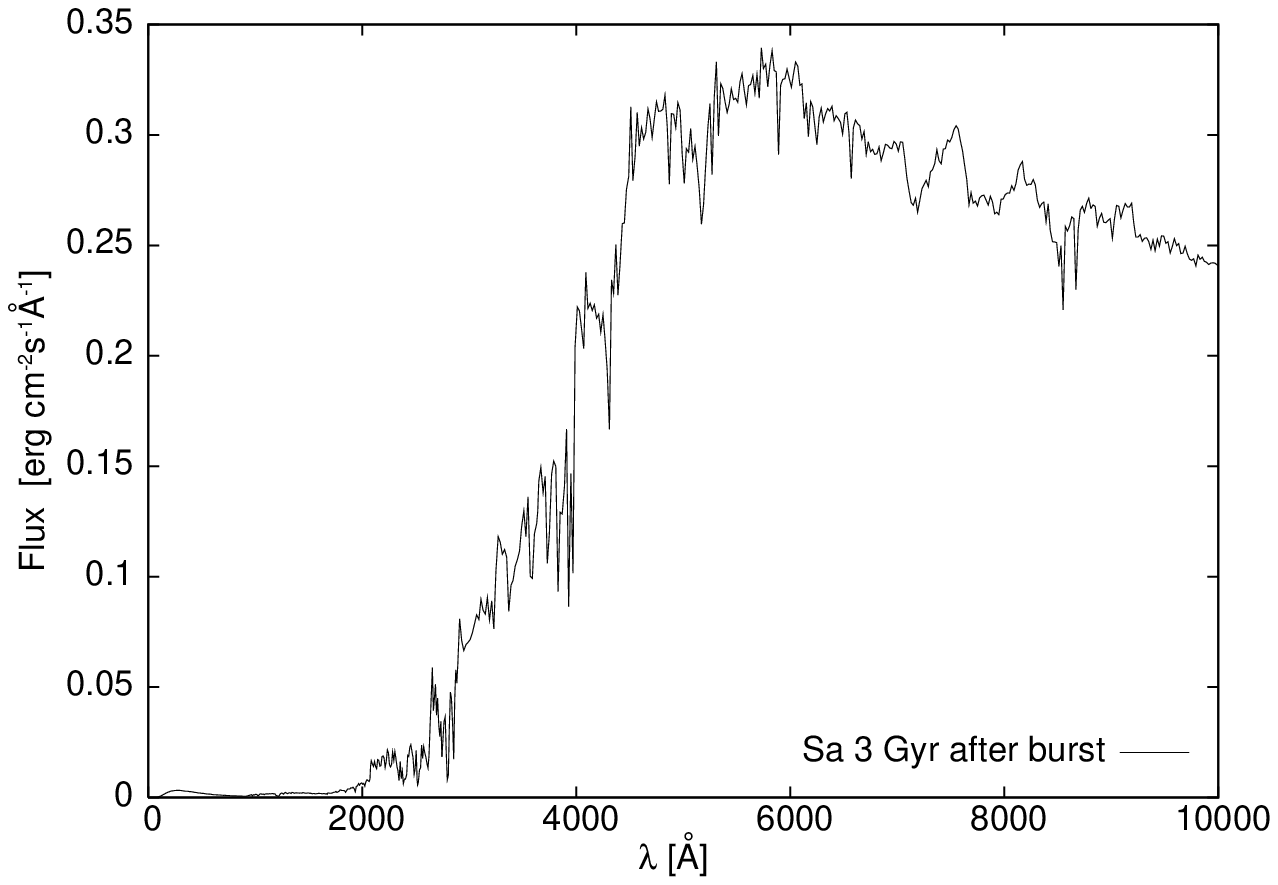}
\includegraphics[width=84mm]{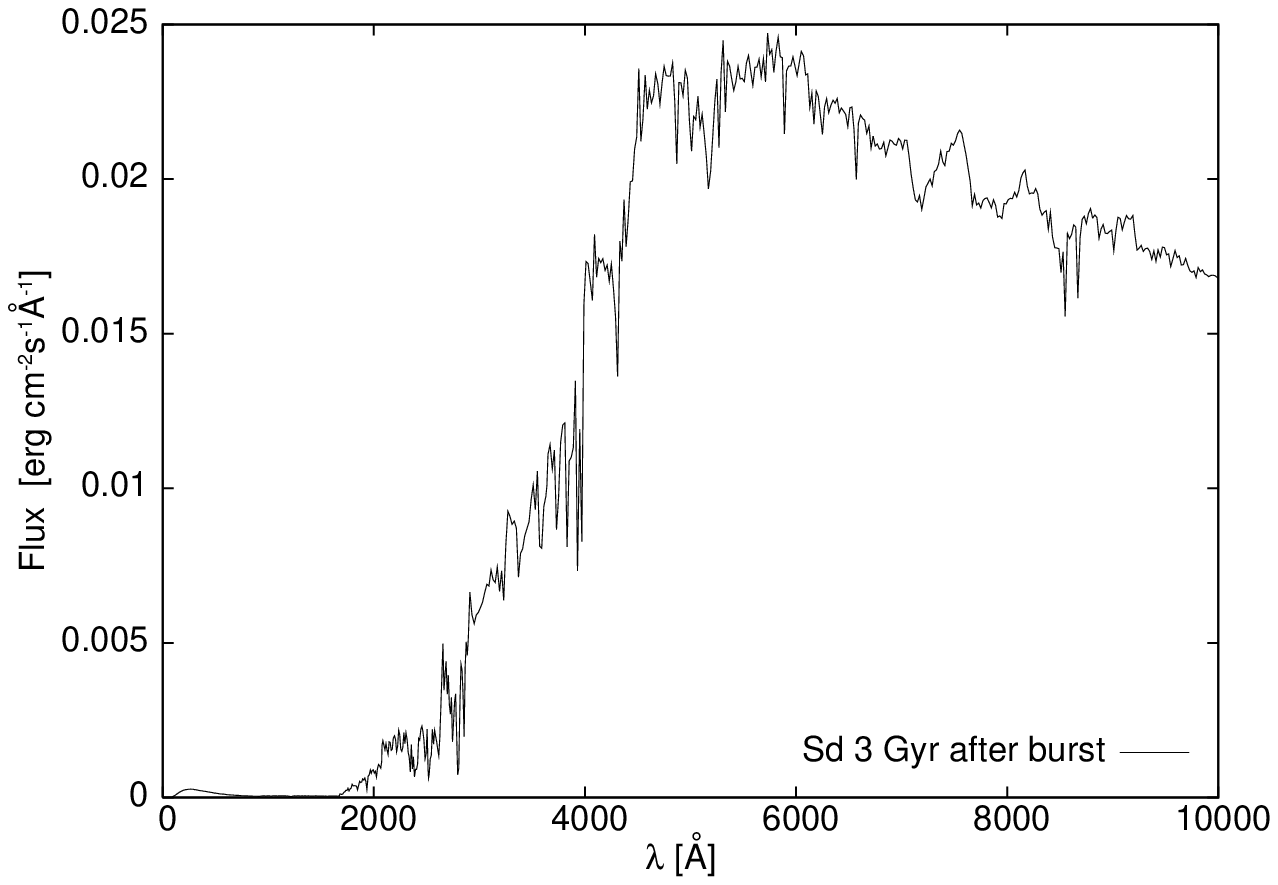}
\caption{Spectra of an Sa model (burst 50\%) and an Sd model (burst 70\%) after their post-starburst phase,  3 Gyr after the bursts (at 6 Gyr, decline times 0.1 Gyr).}
\label{burst_spectra_4}
\end{figure}

Figs. \ref{burst_spectra_1} to \ref{burst_spectra_4} show spectra of the galaxy
models with the highest and the lowest peak value for EW(H$\delta$). The highest
value is reached by the Sd model with a burst strength b=70\%, an onset at 6 Gyr
and a decline time of $\tau$=0.1 Gyr (see Sect. \ref{Hd}), shown on the top of
Figs. \ref{burst_spectra_1} to \ref{burst_spectra_4}. The bottom of
Figs. \ref{burst_spectra_1} to \ref{burst_spectra_4} shows the Sa model with a
burst strength b=50\%, an onset at 6 Gyr and a decline time $\tau$=0.1 Gyr,
which has the lowest value for EW(H$\delta$) of all our galaxy models classified
as E+A. Note the different flux scales in all graphs of
Figs. \ref{burst_spectra_1} to \ref{burst_spectra_4}.  The different stages of
the galaxies in these figures are given in relation to the onset of the burst.
The two graphs in Fig. \ref{burst_spectra_1} show the spectra of the Sd and Sa
burst models at an age of 5 Gyr, i.e. 1 Gyr before the burst starts. Therefore
the galaxies have spectra of normal undisturbed Hubble type Sd and Sa galaxies
at that age.  The graphs in Fig. \ref{burst_spectra_2} show the same galaxies
during their bursts at an age of 6 Gyr. The spectra show much stronger emission
lines and a higher UV flux than the undisturbed galaxies.

In Fig. \ref{burst_spectra_3} the H$\delta$-strong phase of the two models is
shown. The spectra show the galaxies at the age, when the EW(H$\delta$) has
reached its maximum value (i.e. at 6.6 Gyr for the Sd, and at 6.4 Gyr for the Sa
model) and at an age of 7 Gyr, i.e. 1 Gyr after the burst. In
Tab. \ref{Hd_[OII]} the corresponding line strengths for EW(H$\delta$) and
EW([\mbox{O\,{\sc ii}}]) are listed for these spectra. We recall that our
spectra do not have enough resolution to measure the line strength. The values
for the line strengths of EW(H$\delta$) and EW([\mbox{O\,{\sc ii}}]) listed here
are calculated as described in Sect. \ref{OII_rel} and \ref{lick}.

\begin{table}
\caption{Values of EW(H$\delta$) and EW([\mbox{O\,{\sc ii}}]) for an Sd model
  (top) with a burst strength of 70\% and an Sa model (bottom) with burst
  strength of 50\%, both with an onset at 6 Gyr and a decline timescale of 0.1
  Gyr.}
\label{Hd_[OII]}
\centering
\begin{tabular}{l l l l}\hline \hline
Sd & Age & EW(H$\delta$)[\AA]   & EW([\mbox{O\,{\sc ii}}])[\AA]  \\ \hline 
& 6.6 & 8.5 & 4.9 \\
& 7.0 & 7.1 & 0.2\\
& 9.0 & 1.3 & - \\ \hline
\end{tabular}
\begin{tabular}{l l l l}\hline \hline 
Sa & Age & EW(H$\delta$)[\AA]   & EW([\mbox{O\,{\sc ii}}])[\AA] \\ \hline 
& 6.4 & 5.6& 11.7  \\
& 7.0 & 3.4 & 0.1\\
& 9.0 & 0.3 & - \\ \hline
\end{tabular}
\end{table}

The spectra at 6.6 Gyr and 6.4 Gyr, respectively, still have high UV fluxes and
the overall continuum does not yet look like a typical passive galaxy spectrum
but instead resembles an A-star spectrum. In comparison, the two spectra look
rather different, especially when looking at the regions around the
Lyman-$\alpha$-line at 1216 \AA\ and the 4000 \AA\ break.  The Balmer lines are
stronger in the Sd galaxy and no significant emission lines can be seen. In
fact, the Sd galaxy at 6.6 Gyr has an EW([\mbox{O\,{\sc ii}}]) lower than 5
\AA\ and would be classified as an E+A galaxy. However, its spectrum at the
bottom of Fig. \ref{burst_spectra_3} does not look like a typical E+A spectrum,
i.e. like a spectrum of a passive galaxy with significant A-star features,
particularly when looking at the UV flux. This confusion is due to the fact that
in the literature, spectra of E+A galaxies do not reach into the UV region,
which, however, is the wavelength range with the strongest changes during the
post-starburst phase.

The Sa model has strong [\mbox{O\,{\sc ii}}] emission lines for a longer time
after the burst, while its values for EW(H$\delta$) are weaker and drop faster
than in the Sd model.  In contrast to the Sd model, the Sa galaxy at 6.4 Gyr
still has an EW([\mbox{O\,{\sc ii}}])$\geq$ 5 \AA, as shown in
Tab. \ref{Hd_[OII]}. Its spectrum 1 Gyr after the burst looks like a typical
passive spectrum with a superposition of A-star features. The UV flux is very
low and the Lyman-$\alpha$-line at 1216 \AA\ has disappeared. No emission lines
can be seen, while the Balmer lines and the 4000 \AA\ break are present, even
though EW(H$\delta$) $\leq$ 5 \AA\ for the Sa galaxy model (see
Tab. \ref{Hd_[OII]}). From the continuum ratios as well as from the strengths of
the Balmer lines and the 4000 \AA\ break the different burst scenarios are
distinguishable.

The spectra in Fig. \ref{burst_spectra_4} show the galaxy models after the E+A
phase, at a galaxy age of 9 Gyr. Now, 3 Gyr after the burst both spectra look
like true passive galaxy spectra and the Balmer lines have disappeared.
\emph{We conclude that studying E+A galaxies over a larger wavelength range,
  particularly including the UV range, promises a better discrimination of
  progenitor galaxies and transformation scenarios than only looking at
  EW(H$\delta$) and EW([\mbox{O\,{\sc ii}}]).}  Our results also indicate that
the analysis of Spectral Energy Distributions (SEDs) extending from UV/U trough
optical and eventually NIR passbands will allow to encompass all post-starburst
phases, even those, which are excluded from the classical E+A definition based
on the optical region only. This will be investigated Paper II.

\subsection{Comparison with Template Spectra}\label{temps}
\begin{figure}
 
\includegraphics[width=84mm]{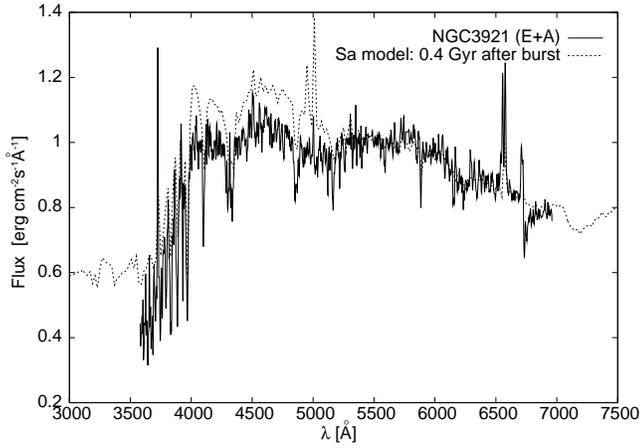}
\caption{Spectra of an Sa model shown 0.4 Gyr after a burst and an E+A template from \citet{Kennicutt1992ApJS} (Sa, burst 50\% at 6 Gyr, decline time 0.1 Gyr).}
\label{kennicutt_Sa_EA_0.4}
 
\end{figure}

\begin{figure}
 
\includegraphics[width=84mm]{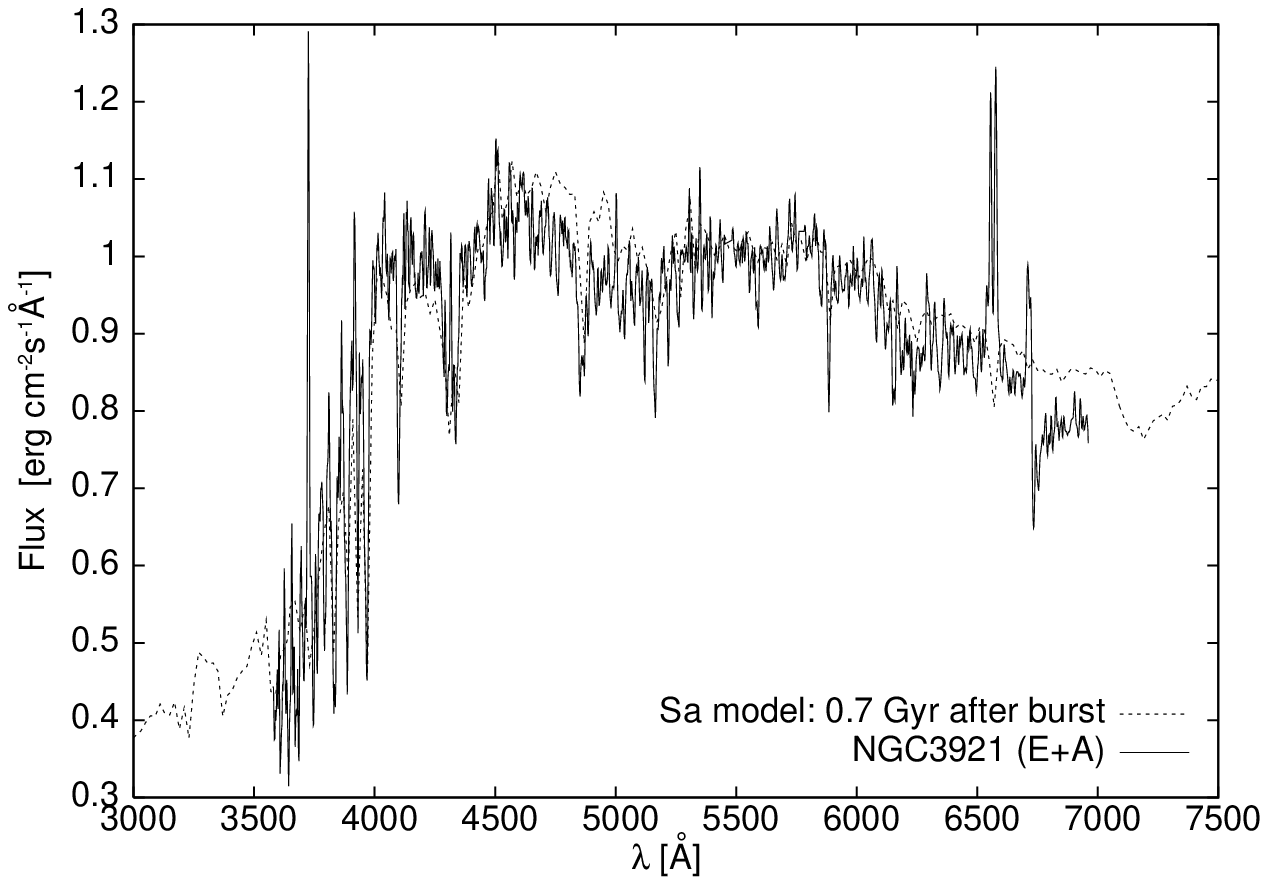}
\caption{Spectra of an Sa model shown 0.7 Gyr after a burst and an E+A template from \citet{Kennicutt1992ApJS} (burst 50\% at 6 Gyr, decline time 0.1 Gyr).}
\label{kennicutt_Sa_EA_0.7}
 
\end{figure}

\begin{figure}
 
\includegraphics[width=84mm]{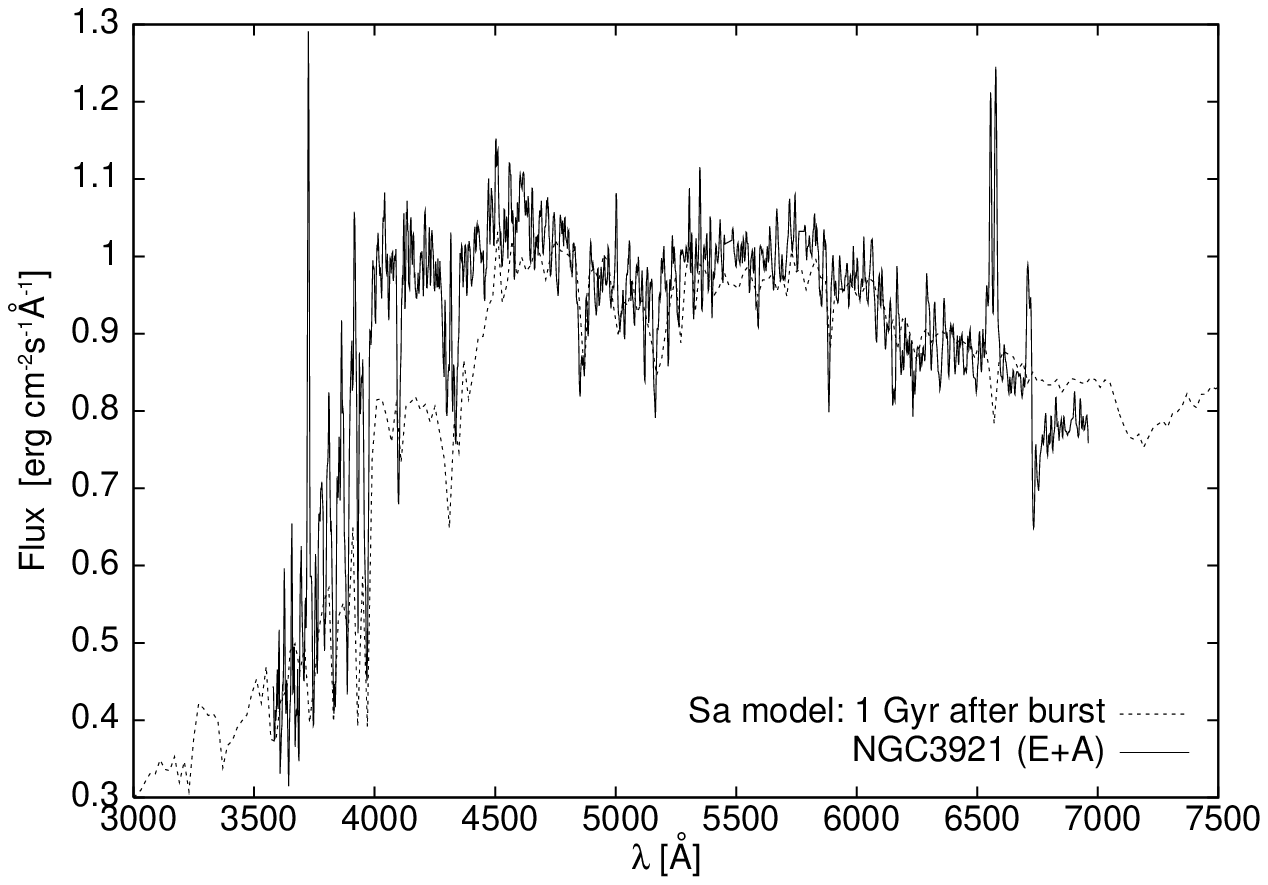}
\caption{Spectra of an Sa model shown 1 Gyr after a burst and an E+A template from \citet{Kennicutt1992ApJS} (burst 50\% at 6 Gyr, decline time 0.1 Gyr).}
\label{kennicutt_Sa_EA_1}
 
\end{figure}

\begin{figure}
 
\includegraphics[width=58mm, angle=270]{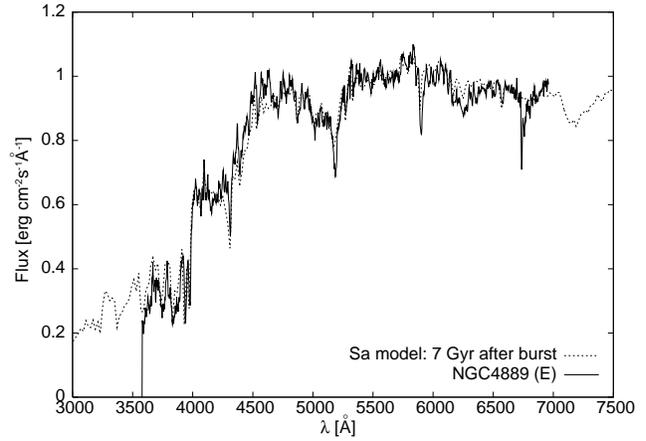}
\caption{Spectra of an Sa model shown 7 Gyr after a burst and an E template from \citet{Kennicutt1992ApJS} (burst 50\% at 6 Gyr, decline time 0.1 Gyr).}
\label{kennicutt_Sa_E}
 
\end{figure}

In Figs. \ref{kennicutt_Sa_EA_0.4} to \ref{kennicutt_Sa_EA_1}, different ages of
the Sa galaxy model with a burst onset at 6 Gyr, a burst strength of 50\% and a
decline time of 0.1 Gyr are shown together with the E+A galaxy template from
\citet{Kennicutt1992ApJS}.  Fig. \ref{kennicutt_Sa_EA_0.4} shows the Sa model
0.4 Gyr after the burst. At this age, the model spectrum is still too young to
match the template. The flux between 4000 and 5000 \AA\ still is too high. When
we look at Fig. \ref{kennicutt_Sa_EA_0.7}, we can see that 0.7 Gyr after the
burst the flux decreased enough to match the template very well. A little later,
1 Gyr after the burst, the flux between 4000 and 5000 \AA\ is already too low as
compared to the template continuum.  Fig. \ref{kennicutt_Sa_E} shows the same Sa
galaxy model 7 Gyr after the burst, compared to Kennicut's E galaxy
template. The model has passed the E+A phase and matches the template of an E
galaxy very well.

\begin{figure}
 
\includegraphics[width=84mm]{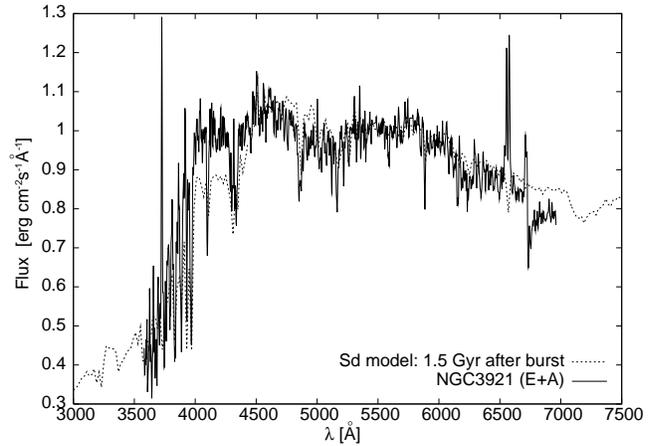}
\caption{Spectra of an Sd model shown 1.5 Gyr after a burst and an E+A template from \citet{Kennicutt1992ApJS} (burst 70\% at 6 Gyr, decline time 0.1 Gyr).}
\label{kennicutt_Sd_EA}
 
\end{figure}

\begin{figure}
 
\includegraphics[width=84mm]{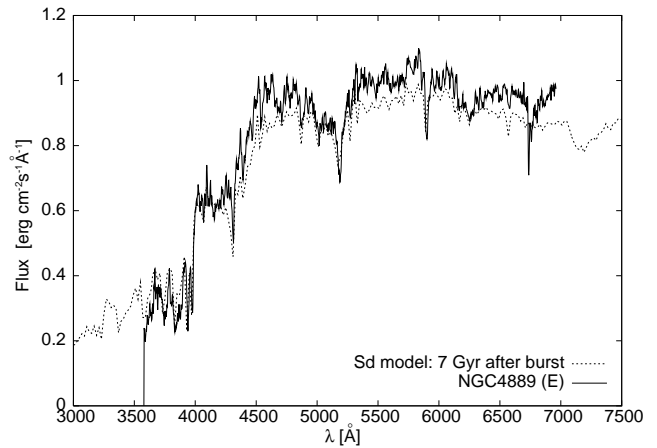}
\caption{Spectra of an Sd model shown 7 Gyr after a burst and an E template from \citet{Kennicutt1992ApJS} (burst 70\% at 6 Gyr, decline time 0.1 Gyr).}
\label{kennicutt_Sd_E}
 
\end{figure}

In Figs. \ref{kennicutt_Sd_EA} and \ref{kennicutt_Sd_E}, we compare the Sd model
with a burst of 70\%, an onset at 6 Gyr and a decline time of 0.1 Gyr with
Kennicut's E+A and E templates, respectively. Shown are the Sd model spectra at
the times when they best match.

It is obvious that, in general, the model spectra match the templates very well,
however not quite as well as the Sa model spectra. This result is inline with
investigations of NGC 3\,921 in the literature. It is proposed that NGC 3921 is
a merger remnant of two disk galaxies, Sa and Sc, with a burst 0.7+/$-$0.3 Gyr
ago, which is expected to evolve into a normal elliptical galaxy on a timescale
of 

Our comparison of E+A galaxy spectra several Gyr after the starburst with template spectra of E and S0 galaxies shows that E+A evolve into early-type galaxies in agreement with investigations from \citet{Bicker2003Ap&SS}  of the origin of S0 galaxies in clusters.

\subsection{B$-$R Color of Post-Starburst Galaxies}\label{chapter_B-R}
\begin{figure}
 
\includegraphics[width=84mm]{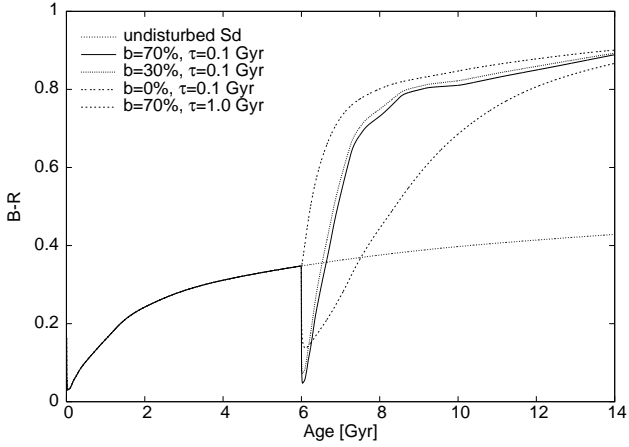}
\caption{Time evolution of B$-$R for Sd models with different burst strengths (Sd, burst beginning at 6 Gyr, decline time 0.1 Gyr).}
\label{B-R_Sd_burst}
 
\end{figure}

In Fig. \ref{B-R_Sd_burst} three Sd models with the same onset of the burst at 6
Gyr, the same decline timescale of $\tau$=0.1 Gyr but different burst strength
(70\%, 30\% and 0\%, i.e. SF truncation) are presented as well as an Sd model
with an onset at 6 Gyr and a decline timescale of $\tau$=1.0 Gyr with a burst
strength of 70\%.  For comparison an undisturbed Sd model is shown.  We recall
that the scenario with the long decline timescale does not lead to an E+A phase
because it shows significant [\mbox{O\,{\sc ii}}]-lines during its
H$\delta$-strong phase.  A burst causes the galaxies to become very blue before
they get red due to the truncation/termination of SF.  The models with SF
truncation or termination only quickly become red immediately after the
onset. Therefore up to 12 Gyr the models with a starbust before the halting of
SF are bluer. Late type spiral galaxies are in general bluer than early type
galaxies and therefore take longer to become red.  For old ages all burst and SF
truncation/termination scenarios converge to similar colors as the age
differences between their stellar populations become more and more negligible
and the photometry of the galaxies is dominated by stars of the Red Giant Branch
and Asymptotic Giant Branch.

\begin{figure}
 
\includegraphics[width=84mm]{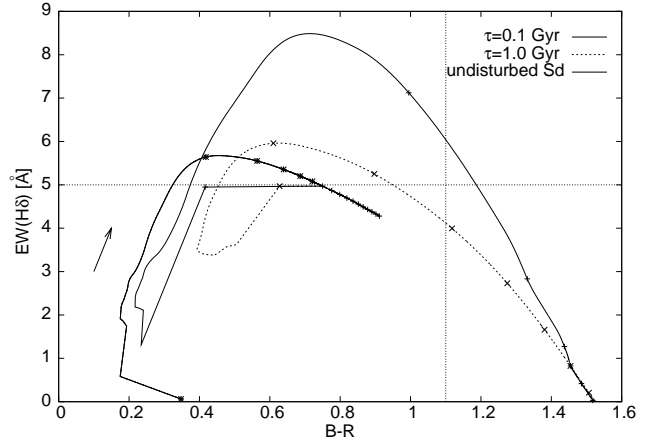}
\caption{EW(H$\delta$) versus B$-$R of two Sd models with different decline times (Sd, burst 70\% at 6 Gyr).}
\label{Hd_B-R_burst}
 
\end{figure}

The decline timescale of the burst has a strong effect on how fast the galaxy
model becomes red after the burst. The model with a long decline time of 1.0 Gyr
is bluer than the undisturbed model for almost 2 Gyr, while the model with the
shortest decline timescale of 0.1 Gyr is already redder than the undisturbed
model after about 0.5 Gyr.  The symbols mark 1 Gyr intervals.  In
Fig. \ref{Hd_B-R_burst} the EW(H$\delta$) is plotted against the color
B$-$R. With this plot it is possible to investigate the evolution in color
during the H$\delta$-strong phase. It is possible to see which galaxy models
become red during their H$\delta$-strong phase and which remain blue. Since blue
as well as red E+A galaxies have been observed this investigation is of great
interest \citep{Yang2008ApJ, Zabludoff1996}.  The horizontal line is the 5
\AA\ threshold, separating H$\delta$-strong from non H$\delta$-strong models,
while the vertical line at B$-$R=1.1 separates blue (to the left) from red
galaxies (on the right).

The value of 1.1 for the limit was chosen to lie between the values of the
undisturbed Sb and Sc models at a galaxy age of 13 Gyr, i.e. at redshift z=0.
Model in the upper right square are red during their H$\delta$-strong phase. The
arrow points into the direction in which the models evolve with time.

In Fig. \ref{Hd_B-R_burst} two Sd models with the same burst strength of 70\%
and an onset at 6 Gyr but different decline times $\tau$ (0.1 and 1.0 Gyr) are
shown together with the undisturbed Sd model. Again it has to be remarked that
the model with the long decline time is not an E+A galaxy.  The burst models
separate from the undisturbed model after the 6 Gyr mark due to the rapid change
in color at the onset of the burst. At the beginning of the burst, the color
starts to change faster than the EW(H$\delta$).  The model with the decline time
of 1.0 Gyr falls already under the 5 \AA\ threshold before it reaches the red
side, i.e. it spends all of its H$\delta$-strong phase in the blue.  The model
with a shorter decline time of 0.1 Gyr crosses the blue-red border while still
H$\delta$-strong.  It is observable both in a blue and in a red H$\delta$-strong
phase because the value for EW(H$\delta$) of the model with decline time 0.1 Gyr
reaches a high peak value for EW(H$\delta$) and drops slowlier below 5 \AA\ than
the color evolves towards the red.  The two models in Fig. \ref{Hd_B-R_burst}
are representative for all models of our grid. Hence \emph{all E+A galaxy models
  with short decline times become red, while models with long decline times (not
  E+A!) always stay blue and do not cross the color threshold.}

For models with a decline time of 0.3 Gyr it depends on the time when the burst
occurs wether they get red during their H$\delta$-strong phase or not.  All
models with a decline time of 0.3 Gyr and an onset at 11 Gyr become red during
their H$\delta$-strong phase, while models with a decline time of 0.3 Gyr and an
onset at 9 and 6 Gyr can show either behavior. Some of these models have
stronger values for EW(H$\delta$) and become red during their H$\delta$-strong
phase, while others have EW(H$\delta$) only marginally above the 5
\AA\ threshold and do not stay H$\delta$-strong long enough to have a red
H$\delta$-strong phase.

We conclude, that the later the type and the longer the decline time, the longer
it takes for the models to become red.  Red post-starburst galaxies are possibly
a later stage of blue post-starburst galaxies with a strong burst, i.e. a merger
or gas stripping scenario in a late stage. Scenarios that terminate the SFR on a
long timescale, like those associated with harassment and strangulation, or
scenarios with an intermediate decline time with a weak burst can be excluded
for red H$\delta$-strong galaxies.

By looking at the colors of H$\delta$-strong galaxies we can constrain their
decline times. This result let us assume that these scenarios are
distinguishable by comparing the SEDs of post-starburst galaxies as investigated
in Paper II. 

\section{Conclusions and Outlook}\label{conclusion}

We investigated the transformation of galaxies when entering a high density
environment like a galaxy cluster. In particular we focused on the E+A galaxy
class, which contains galaxies in an important transition stage, a
post-starburst phase, of galaxy transformation. E+A galaxies are defined to have
strong Balmer absorption lines but no significant emission lines. The spectrum
of an E+A galaxy therefore looks like a superposition of the passive spectrum of
an elliptical galaxy and deep Balmer absorption lines typical for A-stars.  We
examined the different processes of galaxy transformation like mergers,
harassment, gas stripping and strangulation in terms of their impact on the SF
history of galaxies and identified the most likely progenitors and successors of
E+A galaxies.  For this purpose, we used the GALEV galaxy evolution code to
calculate a grid of spiral galaxy models with different combinations of a
starburst, i.e. a sudden increase of the star formation rate (SFR), and a
halting of the SFR on different timescales, i.e. SF truncation and
termination. With these scenarios it is possible to describe the effects on the
SFR of the four different galaxy transformation scenarios mentioned above. We
varied the amount of gas consumed in the starburst (burst strength) and the
galaxy age at which the starburst, the SF truncation or termination begins
(onset).  The spectra and colors as well as the time evolution of Lick indices,
in particular the EW(H$\delta$) of the H$\delta$ absorption line at 4100
\AA\ and the EW([\mbox{O\,{\sc ii}}]) of the [\mbox{O\,{\sc ii}}] emission line
at 3727 \AA\ as indicators of recent and ongoing SF were modelled and discussed.

When investigating the time evolution of the EW(H$\delta$), we found that the
EW(H$\delta$) is the result of a complex interplay of burst strength, decline
timescale, onset of the burst and progenitor galaxy type:

\begin{enumerate}
\item The later the \emph{progenitor galaxy type}, the easier an
  H$\delta$-strong phase is reached. All Sd burst models become H$\delta$-strong
  and the maximum value for EW(H$\delta$) is reached by an Sd model.  For a
  certain type of progenitor spiral galaxy, the essential parameter to cause an
  EW(H$\delta$) above 5 \AA\ is the peak burst SFR $\psi_\mathrm{burst}$ with
  different values for each spiral type.

\item The earlier the \emph{onset} of the burst, the higher are the peak values
  reached by the EW(H$\delta$). For bursts beginning at a galaxy age of 3 Gyr
  all scenarios are H$\delta$-strong due to the larger available gas reservoir
  at such young ages.

\item The longer the \emph{decline timescale} of the SFR after a burst, the
  lower are the values reached for EW(H$\delta$), but the longer is the duration
  of the H$\delta$-strong phase. The H$\delta$- strong phase can last up to 2
  Gyr in cases of long decline timescales.

\item The higher the \emph{burst strength}, the higher are the peak values for
  the EW(H$\delta$). Only very few scenarios with a truncation or termination of
  SF without a preceding burst have a H$\delta$-strong phase. Those are the
  late-type spirals with their relatively high SFRs and a truncation on a short
  timescale. They have EW(H$\delta$) slightly above 5 \AA\ for a short time.
\end{enumerate}

We find a significant effect of the metallicities in galaxies on their
EW(H$\delta$) and conclude that a higher number of E+A galaxies at high redshift
cannot be ascribed solely to a higher number of starburst galaxies but also to
the lower metallicities of galaxies that cause longer H$\delta$-strong phases in
the early universe and to the higher burst strengths enabled by the higher gas
content.

A striking result are the values for EW(H$\delta$) that we found for undisturbed
galaxies. We discovered that galaxies, which only follow the normal evolution,
all go through a phase between ages of 1.5 and 6 Gyr in which they have strong
H$\delta$-lines. Not taking this result into account in analyses of high
redshift galaxies leads to an underestimation of emission lines, which are
influenced by the Balmer absorption lines, for example the $H{\alpha}$-line and
$H{\beta}$, and of the SFR derived from those.

By investigating the second criterion for E+A galaxies, the [\mbox{O\,{\sc ii}}]
emission line at 3727 \AA, we found that the selection criteria for E+A galaxies
commonly used exclude a significant number of post-starburst galaxies, in
particular those galaxies with a long decline time.  The conventional definition
of an E+A galaxy is too narrow to encompass the full range of post-starburst
galaxies.  Possible progenitors of E+A galaxies can only be galaxy mergers and
maybe galaxies with gas stripping with a starburst, while the processes of
harassment and strangulation can be excluded due to the long decline timescales
they induce for SF.

The comparison of E+A galaxy spectra several Gyr after the starburst with
template spectra of E and S0 galaxies showed that the successors of E+A galaxies
are early-type galaxies.  From our model spectra of E+A galaxies, we predict
that studying E+A galaxies over a larger wavelength range, particularly
including the UV, yields a better discrimination of progenitor galaxies and
transformation scenarios. The analysis of Spectral Energy Distributions (SEDs)
will allow to encompass all post-starburst phases, even those which are excluded
from the classical E+A definition based on optical regions only.  From our
investigation of the time evolution of the B$-$R colors of our post-starburst
galaxy models we found, that the progenitor galaxy type and the decline time
have the greatest effects on the color evolution. The later the type and the
longer the decline time, the longer it takes for the models to become red.  Red
E+A galaxies and red post-starburst galaxies in general are possibly a later
stage of blue post-starburst galaxies with a strong burst and a short or
intermediate decline time as e.g. triggered by a merger. Galaxy models with an
intermediate decline time ($\geq$0.3 Gyr) with a weak burst and models with a
long decline time ($\geq$1.0 Gyr) no matter how strong a burst was do not become
red during their H$\delta$-strong phase.  Our results from examining the spectra
and the colors of post-starburst galaxies indicate that it is possible to select
and distinguish post-starburst galaxies by looking at their Spectral Energy
Distributions (SEDs), which will be presented in Paper II.

We have shown that E+A galaxies by definition include only transformation
processes on a short timescale. The full range of post-starburst galaxies have
not yet been investigated observationally. This needs to be done in order to
achieve a comprehensive understanding of the galaxy transformation processes in
high density environments.

\section*{Acknowledgments}
We would like to thank the referee for very useful comments and suggestions.

 
\bibliographystyle{mn2e}
\bibliography{paper1}

\begin{thebibliography}{}

\bibitem[\protect\citeauthoryear{{Anders} \& {Fritze-v.~Alvensleben}}{{Anders}
  \& {Fritze-v.~Alvensleben}}{2003}]{Anders2003A&A}
{Anders} P.,  {Fritze-v.~Alvensleben} U.,  2003, \aap, 401, 1063

\bibitem[\protect\citeauthoryear{{Balogh}, {Miller}, {Nichol}, {Zabludoff} \&
  {Goto}}{{Balogh} et~al.}{2005}]{Balogh2005MNRAS}
{Balogh} M.~L.,  {Miller} C.,  {Nichol} R.,  {Zabludoff} A.,    {Goto} T.,
  2005, \mnras, 360, 587

\bibitem[\protect\citeauthoryear{{Balogh}, {Morris}, {Yee}, {Carlberg} \&
  {Ellingson}}{{Balogh} et~al.}{1999}]{Balogh1999}
{Balogh} M.~L.,  {Morris} S.~L.,  {Yee} H.~K.~C.,  {Carlberg} R.~G.,
  {Ellingson} E.,  1999, \apj, 527, 54

\bibitem[\protect\citeauthoryear{{Barger}, {Aragon-Salamanca}, {Ellis},
  {Couch}, {Smail} \& {Sharples}}{{Barger} et~al.}{1996}]{Barger1996MNRAS}
{Barger} A.~J.,  {Aragon-Salamanca} A.,  {Ellis} R.~S.,  {Couch} W.~J.,
  {Smail} I.,    {Sharples} R.~M.,  1996, \mnras, 279, 1

\bibitem[\protect\citeauthoryear{{Bekki}, {Couch}, {Shioya} \&
  {Vazdekis}}{{Bekki} et~al.}{2005}]{Bekki2005MNRAS}
{Bekki} K.,  {Couch} W.~J.,  {Shioya} Y.,    {Vazdekis} A.,  2005, \mnras, 359,
  949

\bibitem[\protect\citeauthoryear{{Bertelli}, {Bressan}, {Chiosi}, {Fagotto} \&
  {Nasi}}{{Bertelli} et~al.}{1994}]{Bertelli1994A&AS}
{Bertelli} G.,  {Bressan} A.,  {Chiosi} C.,  {Fagotto} F.,    {Nasi} E.,  1994,
  \aaps, 106, 275

\bibitem[\protect\citeauthoryear{{Bicker}, {Fritze-v.~Alvensleben},
  {M{\"o}ller} \& {Fricke}}{{Bicker} et~al.}{2004}]{Bicker2004A&A}
{Bicker} J.,  {Fritze-v.~Alvensleben} U.,  {M{\"o}ller} C.~S.,    {Fricke}
  K.~J.,  2004, \aap, 413, 37

\bibitem[\protect\citeauthoryear{{Bicker}, {Fritze-von Alvensleben} \&
  {Fricke}}{{Bicker} et~al.}{2003}]{Bicker2003Ap&SS}
{Bicker} J.,  {Fritze-von Alvensleben} U.,    {Fricke} K.~J.,  2003, \apss,
  284, 463

\bibitem[\protect\citeauthoryear{{Bravo-Alfaro}, {Cayatte}, {van Gorkom} \&
  {Balkowski}}{{Bravo-Alfaro} et~al.}{2000}]{Bravo-Alfaro2000AJ}
{Bravo-Alfaro} H.,  {Cayatte} V.,  {van Gorkom} J.~H.,    {Balkowski} C.,
  2000, \aj, 119, 580

\bibitem[\protect\citeauthoryear{{Butcher} \& {Oemler} Jr.}{{Butcher} \&
  {Oemler}}{1978}]{Butcher1978ApJ}
{Butcher} H.,  {Oemler} Jr. A.,  1978, \apj, 226, 559

\bibitem[\protect\citeauthoryear{{Butcher} \& {Oemler} Jr.}{{Butcher} \&
  {Oemler}}{1984}]{Butcher1984}
{Butcher} H.,  {Oemler} Jr. A.,  1984, \apj, 285, 426

\bibitem[\protect\citeauthoryear{{Buyle}, {De Rijcke} \& {Dejonghe}}{{Buyle}
  et~al.}{2008}]{Buyle2008arxiv}
{Buyle} P.,  {De Rijcke} S.,    {Dejonghe} H.,  2008, ArXiv:0807.0440

\bibitem[\protect\citeauthoryear{{Cayatte}, {van Gorkom}, {Balkowski} \&
  {Kotanyi}}{{Cayatte} et~al.}{1990}]{Cayatte1990AJ}
{Cayatte} V.,  {van Gorkom} J.~H.,  {Balkowski} C.,    {Kotanyi} C.,  1990,
  \aj, 100, 604

\bibitem[\protect\citeauthoryear{{Chang}, {van Gorkom}, {Zabludoff}, {Zaritsky}
  \& {Mihos}}{{Chang} et~al.}{2001}]{Chang2001AJ}
{Chang} T.-C.,  {van Gorkom} J.~H.,  {Zabludoff} A.~I.,  {Zaritsky} D.,
  {Mihos} J.~C.,  2001, \aj, 121, 1965

\bibitem[\protect\citeauthoryear{{Couch}, {Ellis}, {Sharples} \&
  {Smail}}{{Couch} et~al.}{1994}]{Couch1994ApJ}
{Couch} W.~J.,  {Ellis} R.~S.,  {Sharples} R.~M.,    {Smail} I.,  1994, \apj,
  430, 121

\bibitem[\protect\citeauthoryear{{Couch} \& {Sharples}}{{Couch} \&
  {Sharples}}{1987}]{Couch1987MNRAS}
{Couch} W.~J.,  {Sharples} R.~M.,  1987, \mnras, 229, 423

\bibitem[\protect\citeauthoryear{{Dressler}}{{Dressler}}{1980}]{Dressler1980Ap%
J}
{Dressler} A.,  1980, \apj, 236, 351

\bibitem[\protect\citeauthoryear{{Dressler}, {Oemler}, {Butcher} \&
  {Gunn}}{{Dressler} et~al.}{1994}]{Dressler1994ApJ}
{Dressler} A.,  {Oemler} A.~J.,  {Butcher} H.~R.,    {Gunn} J.~E.,  1994, \apj,
  430, 107

\bibitem[\protect\citeauthoryear{{Dressler}, {Oemler}, {Couch}, {Smail},
  {Ellis}, {Barger}, {Butcher}, {Poggianti} \& {Sharples}}{{Dressler}
  et~al.}{1997}]{Dressler1997}
{Dressler} A.,  {Oemler} A.~J.,  {Couch} W.~J.,  {Smail} I.,  {Ellis} R.~S.,
  {Barger} A.,  {Butcher} H.,  {Poggianti} B.~M.,    {Sharples} R.~M.,  1997,
  \apj, 490, 577

\bibitem[\protect\citeauthoryear{{Dressler}, {Smail}, {Poggianti}, {Butcher},
  {Couch}, {Ellis} \& {Oemler}}{{Dressler} et~al.}{1999}]{Dressler1999}
{Dressler} A.,  {Smail} I.,  {Poggianti} B.~M.,  {Butcher} H.,  {Couch} W.~J.,
  {Ellis} R.~S.,    {Oemler} A.~J.,  1999, \apjs, 122, 51

\bibitem[\protect\citeauthoryear{{Ellingson}, {Lin}, {Yee} \&
  {Carlberg}}{{Ellingson} et~al.}{2001}]{Ellingson2001ApJ}
{Ellingson} E.,  {Lin} H.,  {Yee} H.~K.~C.,    {Carlberg} R.~G.,  2001, \apj,
  547, 609

\bibitem[\protect\citeauthoryear{{Falkenberg}, {Kotulla} \&
  {Fritze}}{{Falkenberg} et~al.}{2009}]{Falkenberg+09b}
{Falkenberg} M.~A.,  {Kotulla} R.,    {Fritze} U.,  2009, submitted to MNRAS

\bibitem[\protect\citeauthoryear{{Fasano}, {Poggianti}, {Couch}, {Bettoni},
  {Kj{\ae}rgaard} \& {Moles}}{{Fasano} et~al.}{2000}]{Fasano2000}
{Fasano} G.,  {Poggianti} B.~M.,  {Couch} W.~J.,  {Bettoni} D.,
  {Kj{\ae}rgaard} P.,    {Moles} M.,  2000, \apj, 542, 673

\bibitem[\protect\citeauthoryear{{Gavazzi}, {Cortese}, {Boselli},
  {Iglesias-Paramo}, {V{\'{\i}}lchez} \& {Carrasco}}{{Gavazzi}
  et~al.}{2003}]{Gavazzi2003ApJ}
{Gavazzi} G.,  {Cortese} L.,  {Boselli} A.,  {Iglesias-Paramo} J.,
  {V{\'{\i}}lchez} J.~M.,    {Carrasco} L.,  2003, \apj, 597, 210

\bibitem[\protect\citeauthoryear{{Girardi}, {Bertelli}, {Bressan}, {Chiosi},
  {Groenewegen}, {Marigo}, {Salasnich} \& {Weiss}}{{Girardi}
  et~al.}{2003}]{Girardi2003MmSAI}
{Girardi} L.,  {Bertelli} G.,  {Bressan} A.,  {Chiosi} C.,  {Groenewegen}
  M.~A.~T.,  {Marigo} P.,  {Salasnich} B.,    {Weiss} A.,  2003, Memorie della
  Societa Astronomica Italiana, 74, 474

\bibitem[\protect\citeauthoryear{{G{\'o}mez}, {Nichol}, {Miller}, {Balogh},
  {Goto}, {Zabludoff}, {Romer}, {Bernardi}, {Sheth}, {Hopkins}, {Castander},
  {Connolly}, {Schneider}, {Brinkmann}, {Lamb}, {SubbaRao} \&
  {York}}{{G{\'o}mez} et~al.}{2003}]{Goto2003ApJ}
{G{\'o}mez} P.~L.,  {Nichol} R.~C.,  {Miller} C.~J.,  {Balogh} M.~L.,  {Goto}
  T.,  {Zabludoff} A.~I.,  {Romer} A.~K.,  {Bernardi} M.,  {Sheth} R.,
  {Hopkins} A.~M.,  {Castander} F.~J.,  {Connolly} A.~J.,  {Schneider} D.~P.,
  {Brinkmann} J.,  {Lamb} D.~Q.,  {SubbaRao} M.,    {York} D.~G.,  2003, \apj,
  584, 210

\bibitem[\protect\citeauthoryear{{Goto}}{{Goto}}{2004}]{Goto2004A&A}
{Goto} T.,  2004, \aap, 427, 125

\bibitem[\protect\citeauthoryear{{Gunn} \& {Gott}}{{Gunn} \&
  {Gott}}{1972}]{Gunn1972ApJ}
{Gunn} J.~E.,  {Gott} J.~R.~I.,  1972, \apj, 176, 1

\bibitem[\protect\citeauthoryear{{Helmboldt}}{{Helmboldt}}{2007}]{Helmboldt200%
7MNRAS}
{Helmboldt} J.~F.,  2007, \mnras, 379, 1227

\bibitem[\protect\citeauthoryear{{Kaviraj}, {Kirkby}, {Silk} \&
  {Sarzi}}{{Kaviraj} et~al.}{2007}]{Kavirai2007MNRAS}
{Kaviraj} S.,  {Kirkby} L.~A.,  {Silk} J.,    {Sarzi} M.,  2007, \mnras, 382,
  960

\bibitem[\protect\citeauthoryear{{Kennicutt}
  Jr.}{{Kennicutt}}{1992a}]{Kennicutt1992ApJS}
{Kennicutt} Jr. R.~C.,  1992a, \apjs, 79, 255

\bibitem[\protect\citeauthoryear{{Kennicutt}
  Jr.}{{Kennicutt}}{1992b}]{Kennicutt1992ApJ}
{Kennicutt} Jr. R.~C.,  1992b, \apj, 388, 310

\bibitem[\protect\citeauthoryear{{Kodama} \& {Bower}}{{Kodama} \&
  {Bower}}{2001}]{Kodama2001MNRAS}
{Kodama} T.,  {Bower} R.~G.,  2001, \mnras, 321, 18

\bibitem[\protect\citeauthoryear{{Lejeune}, {Cuisinier} \& {Buser}}{{Lejeune}
  et~al.}{1998}]{Lejeune1998A&AS}
{Lejeune} T.,  {Cuisinier} F.,    {Buser} R.,  1998, \aaps, 130, 65

\bibitem[\protect\citeauthoryear{{Leonardi} \& {Rose}}{{Leonardi} \&
  {Rose}}{1996}]{Leonardi1996AJ}
{Leonardi} A.~J.,  {Rose} J.~A.,  1996, \aj, 111, 182

\bibitem[\protect\citeauthoryear{{Lilly} \& {Fritze}}{{Lilly} \&
  {Fritze}}{2006}]{Lilly2006A&A}
{Lilly} T.,  {Fritze} U.~F.-V.,  2006, \aap, 457, 467

\bibitem[\protect\citeauthoryear{{Liu}, {Hooper}, {O'Neil}, {Thompson}, {Wolf}
  \& {Lisker}}{{Liu} et~al.}{2007}]{Liu2007ApJ}
{Liu} C.~T.,  {Hooper} E.~J.,  {O'Neil} K.,  {Thompson} D.,  {Wolf} M.,
  {Lisker} T.,  2007, \apj, 658, 249

\bibitem[\protect\citeauthoryear{{Mihos}}{{Mihos}}{2004}]{Mihos2004cgpc}
{Mihos} J.~C.,  2004, in {Mulchaey} J.~S.,  {Dressler} A.,   {Oemler} A.,  eds,
  Clusters of Galaxies: Probes of Cosmological Structure and Galaxy Evolution
  {Interactions and Mergers of Cluster Galaxies}.
pp 277--+

\bibitem[\protect\citeauthoryear{{Moore}, {Lake} \& {Katz}}{{Moore}
  et~al.}{1998}]{Moore1998ApJ}
{Moore} B.,  {Lake} G.,    {Katz} N.,  1998, \apj, 495, 139

\bibitem[\protect\citeauthoryear{{Oemler}}{{Oemler}}{1974}]{Oemler1974ApJ}
{Oemler} A.~J.,  1974, \apj, 194, 1

\bibitem[\protect\citeauthoryear{{Poggianti}}{{Poggianti}}{2004}]{Poggianti200%
4PoS}
{Poggianti} B.,  2004, in {Dettmar} R.,  {Klein} U.,   {Salucci} P.,  eds,
  Baryons in Dark Matter Halos {Evolution of galaxies in clusters}

\bibitem[\protect\citeauthoryear{{Poggianti} \& {Barbaro}}{{Poggianti} \&
  {Barbaro}}{1996}]{Poggianti1996}
{Poggianti} B.~M.,  {Barbaro} G.,  1996, \aap, 314, 379

\bibitem[\protect\citeauthoryear{{Poggianti}, {Bridges}, {Komiyama}, {Yagi},
  {Carter}, {Mobasher}, {Okamura} \& {Kashikawa}}{{Poggianti}
  et~al.}{2004}]{Poggianti2004ApJ}
{Poggianti} B.~M.,  {Bridges} T.~J.,  {Komiyama} Y.,  {Yagi} M.,  {Carter} D.,
  {Mobasher} B.,  {Okamura} S.,    {Kashikawa} N.,  2004, \apj, 601, 197

\bibitem[\protect\citeauthoryear{{Quilis}, {Moore} \& {Bower}}{{Quilis}
  et~al.}{2000}]{Quilis2000Sci}
{Quilis} V.,  {Moore} B.,    {Bower} R.,  2000, Science, 288, 1617

\bibitem[\protect\citeauthoryear{{Richstone}}{{Richstone}}{1976}]{Richstone197%
6ApJ}
{Richstone} D.~O.,  1976, \apj, 204, 642

\bibitem[\protect\citeauthoryear{{Sandage}, {Binggeli} \& {Tammann}}{{Sandage}
  et~al.}{1985}]{Sandage1985AJb}
{Sandage} A.,  {Binggeli} B.,    {Tammann} G.~A.,  1985, \aj, 90, 1759

\bibitem[\protect\citeauthoryear{{Toomre} \& {Toomre}}{{Toomre} \&
  {Toomre}}{1972}]{Toomre1972ApJ}
{Toomre} A.,  {Toomre} J.,  1972, \apj, 178, 623

\bibitem[\protect\citeauthoryear{{Trager}, {Worthey}, {Faber}, {Burstein} \&
  {Gonzalez}}{{Trager} et~al.}{1998}]{Trager1998ApJS}
{Trager} S.~C.,  {Worthey} G.,  {Faber} S.~M.,  {Burstein} D.,    {Gonzalez}
  J.~J.,  1998, \apjs, 116, 1

\bibitem[\protect\citeauthoryear{{Tremonti}, {Heckman}, {Kauffmann},
  {Brinchmann}, {Charlot}, {White}, {Seibert}, {Peng}, {Schlegel}, {Uomoto},
  {Fukugita} \& {Brinkmann}}{{Tremonti} et~al.}{2004}]{Tremonti2004ApJ}
{Tremonti} C.~A.,  {Heckman} T.~M.,  {Kauffmann} G.,  {Brinchmann} J.,
  {Charlot} S.,  {White} S.~D.~M.,  {Seibert} M.,  {Peng} E.~W.,  {Schlegel}
  D.~J.,  {Uomoto} A.,  {Fukugita} M.,    {Brinkmann} J.,  2004, \apj, 613, 898

\bibitem[\protect\citeauthoryear{{Verdugo}, {Ziegler} \& {Gerken}}{{Verdugo}
  et~al.}{2008}]{Verdugo2008A&A}
{Verdugo} M.,  {Ziegler} B.~L.,    {Gerken} B.,  2008, \aap, 486, 9

\bibitem[\protect\citeauthoryear{{Worthey}, {Faber}, {Gonzalez} \&
  {Burstein}}{{Worthey} et~al.}{1994}]{Worthey1994ApJS}
{Worthey} G.,  {Faber} S.~M.,  {Gonzalez} J.~J.,    {Burstein} D.,  1994,
  \apjs, 94, 687

\bibitem[\protect\citeauthoryear{{Worthey} \& {Ottaviani}}{{Worthey} \&
  {Ottaviani}}{1997}]{Worthey1997ApJS}
{Worthey} G.,  {Ottaviani} D.~L.,  1997, \apjs, 111, 377

\bibitem[\protect\citeauthoryear{{Yang}, {Zabludoff}, {Zaritsky}, {Lauer} \&
  {Mihos}}{{Yang} et~al.}{2004}]{Yang2004}
{Yang} Y.,  {Zabludoff} A.~I.,  {Zaritsky} D.,  {Lauer} T.~R.,    {Mihos}
  J.~C.,  2004, \apj, 607, 258

\bibitem[\protect\citeauthoryear{{Yang}, {Zabludoff}, {Zaritsky} \&
  {Mihos}}{{Yang} et~al.}{2008}]{Yang2008ApJ}
{Yang} Y.,  {Zabludoff} A.~I.,  {Zaritsky} D.,    {Mihos} J.~C.,  2008, \apj,
  688, 945

\bibitem[\protect\citeauthoryear{{Zabludoff}, {Zaritsky}, {Lin}, {Tucker},
  {Hashimoto}, {Shectman}, {Oemler} \& {Kirshner}}{{Zabludoff}
  et~al.}{1996}]{Zabludoff1996}
{Zabludoff} A.~I.,  {Zaritsky} D.,  {Lin} H.,  {Tucker} D.,  {Hashimoto} Y.,
  {Shectman} S.~A.,  {Oemler} A.,    {Kirshner} R.~P.,  1996, \apj, 466, 104

\bibitem[\protect\citeauthoryear{{Zaritsky}, {Kennicutt} Jr. \&
  {Huchra}}{{Zaritsky} et~al.}{1994}]{Zaritsky1994ApJ}
{Zaritsky} D.,  {Kennicutt} Jr. R.~C.,    {Huchra} J.~P.,  1994, \apj, 420, 87

\end{thebibliography}
\label{lastpage}
\end{document}